\documentclass[
  aps,
  pra,
  reprint,
  superscriptaddress,
  nofootinbib
]{revtex4-2}

\usepackage{amsmath}
\usepackage{amssymb}
\usepackage{amsfonts}
\usepackage{physics}

\usepackage{graphicx}
\usepackage{xcolor}
\usepackage[normalem]{ulem}

\usepackage{tikz}
\usetikzlibrary{
  arrows.meta,
  backgrounds,
  calc,
  decorations.pathmorphing,
  fit,
  positioning
}
\usepackage{xcolor}
\usepackage{soul}
\usepackage{pgfplots}
\pgfplotsset{compat=1.18}
\usepackage[hidelinks]{hyperref}

\usepackage{pdfcomment}

\definecolor{cExact}{RGB}{27,110,82}
\definecolor{cCond}{RGB}{186,128,24}
\definecolor{cApprox}{RGB}{178,72,42}
\definecolor{cTwist}{RGB}{42,82,150}
\definecolor{cGrey}{RGB}{105,105,115}

\colorlet{syscol}{cyan!60!black}
\colorlet{sysfill}{cyan!8}
\colorlet{envcol}{orange!70!black}
\colorlet{envfill}{orange!10}

\newcommand{\timeord}{\mathcal{T}}

\newcommand{\ckR}{c_{\alpha\beta,k}^R}
\newcommand{\ckI}{c_{\alpha\beta,k}^I}
\newcommand{\gammakR}{\gamma_{\alpha\beta,k}^R}
\newcommand{\gammakI}{\gamma_{\alpha\beta,k}^I}
\newcommand{\Nr}{N_{\alpha\beta}^R}
\newcommand{\Ni}{N_{\alpha\beta}^I}

\begin{document}

\title{
  Spectral Twisting of Collective Dynamics\\
  of Two Qubits Interacting with a Common Bath
}
\title{Spectral Twisting in a Common Bosonic Reservoir: \\Fragility of Two-Qubit Dark-State Protection} 
\author{Fabio Borrelli}
\affiliation{
  Department of Electrical Engineering and Information Technology,
  Universit\`a degli Studi di Napoli Federico II,
  via Claudio 21, Napoli, 80125, Italy
}

\author{Giovanni Miano}
\affiliation{
  Scuola Superiore Meridionale,
  via Mezzocannone 4, Napoli, 80125, Italy
}
\affiliation{
  Department of Electrical Engineering and Information Technology,
  Universit\`a degli Studi di Napoli Federico II,
  via Claudio 21, Napoli, 80125, Italy
}

\author{Carlo Forestiere}
\email[]{carlo.forestiere@unina.it}
\affiliation{
  Department of Electrical Engineering and Information Technology,
  Universit\`a degli Studi di Napoli Federico II,
  via Claudio 21, Napoli, 80125, Italy
}


\begin{abstract}
The interaction of two qubits with a common bosonic reservoir is encoded by a matrix-valued spectral density $\mathbf{J}(\omega)$, whose diagonal entries describe the local spectra, while the off-diagonal entries encode cross-correlations.  Even when \(\mathbf{J}(\omega)\) has rank one, a frequency-independent dark channel need not exist because the family \(\{\mathbf{J}(\omega)\}_{\omega}\) may have a trivial common kernel.  We term the frequency-dependent rotation of the bright and dark directions \textit{spectral twisting} and quantify it through the Fubini--Study speed $\tau(\omega)$ of the bright spectral projector.  We analyze how twisting modifies two-qubit dynamics and quantify the loss of dark-state protection through the leakage \(P_{\mathrm{leak}}(t)\).  Comparisons with untwisted asymmetric reservoirs and rotating-wave dynamics, together with detuned and finite-temperature calculations, distinguish spectral twisting from coupling asymmetry, counter-rotating processes, and thermal absorption.  For resonant qubits tuned to the crossing of the two local spectra $\omega_\times$, the singlet is locally dark at the transition frequency but couples to off-resonant components whose bright directions are rotated. In the weak-twisting regime, the fixed-time leakage scales as
$P_{\mathrm{leak}}(t)\propto[\omega_\times\tau(\omega_\times)]^2$. We test this prediction for mismatched Drude–Lorentz spectra using nonperturbative hierarchical equations of motion generalized to cross-correlated bath forces.  These results provide a geometric framework for dark-state engineering in structured reservoirs.
\end{abstract}

\maketitle

\section{Introduction}
\label{sec:introduction}

The interaction of a quantum system with its environment governs decoherence, dissipation,  thermalization, and the evolution of quantum correlations. In multipartite systems, coupling to a shared reservoir can also produce correlated noise and mediate coherent and dissipative
interactions between the subsystems. Such correlations produce errors beyond independent-noise models and may challenge quantum-error-correction protocols \cite{klesse_quantum_2005,wilen_correlated_2021}. They have been measured in superconducting circuits \cite{von_lupke_two-qubit_2020} and semiconductor spin qubits \cite{yoneda_noise-correlation_2023}. Conversely, structured correlated reservoirs can mediate interactions and generate long-lived entanglement \cite{zou_spatially_2024}, making them also a resource for quantum-state engineering.

A paradigmatic example is provided by two qubits coupled to a common bosonic environment. They exhibit collective superradiant and subradiant channels
\cite{dicke_coherence_1954,lehmberg_radiation_1970,ficek_entangled_2002}. For identical couplings, destructive interference produces a dark state that, if preserved by the system Hamiltonian, supports decoherence-free subspaces and noiseless encodings
\cite{duan_reducing_1998,zanardi_noiseless_1997,lidar_decoherence-free_2003}. Common-bath models have therefore been studied in connection with collective decay, reservoir-induced entanglement, entanglement trapping, and non-Markovian dynamics
\cite{braun_creation_2002,benatti_environment-induced_2003,maniscalco_protecting_2008,mazzola_sudden_2009,ma_entanglement_2012,wang_exact_2013}, including beyond the rotating-wave, Born, and Markov approximations using hierarchical equations of motion (HEOM) 
\cite{ma_entanglement_2012}.

Cross-correlated fluctuations also arise in molecular aggregates and excitonic complexes \cite{abramavicius_exciton_2011,huo_influence_2012,jing_nonperturbative_2015}, waveguide QED \cite{sheremet_waveguide_2023,arranz_regidor_modeling_2021}, circuit-QED architectures coupled through common transmission lines
\cite{filipp_multimode_2011,parra-rodriguez_quantum_2018,borrelli_dynamical_2026}, and nanophotonic systems containing multiple emitters
\cite{gonzalez-tudela_lightmatter_2024,medina_few-mode_2021,miano_modified_2026}. In these settings, propagation, dispersion, and frequency-selective coupling can make both the magnitude and phase of the cross correlations frequency dependent.

These correlations are described by the matrix-valued spectral density
\(\mathbf{J}(\omega)\). Its diagonal elements are the local spectra, whereas its off-diagonal elements are frequency-resolved cross spectra. These quantities can, in principle, be reconstructed using multiqubit noise spectroscopy
\cite{szankowski_spectroscopy_2016,paz-silva_multiqubit_2017}. 

Within a weak-coupling secular GKLS description, the dissipative Kossakowski matrix is determined by $\mathbf J(\omega)$ evaluated at the system Bohr frequencies, whereas off-resonant spectral components enter only through the principal-value integrals defining the Lamb-shift Hamiltonian \cite{lehmberg_radiation_1970,davies_markovian_1974,gorini_completely_1976,lindblad_generators_1976,cattaneo2019local}. Beyond the weak-coupling, Markov, and secular approximations, the reduced dynamics can depend more generally on the variation of $\mathbf J(\omega)$ over a finite spectral interval.

In particular, the eigendirections of \(\mathbf{J}(\omega)\) may vary with frequency, so that different spectral components couple to different linear combinations of system operators. Consequently, a single frequency-independent transformation need not diagonalize the system--bath coupling over the entire spectrum. Reference~\cite{LeDe2026} identified this frequency dependence as the central obstacle to bath decorrelation and developed numerical strategies to address it.

Here, we investigate the physical consequences of the frequency-dependent rotation of the bright and dark eigendirections of \(\mathbf{J}(\omega)\) for two qubits coupled to a maximally cross-spectrally coherent bosonic environment. For a positive-semidefinite \(2\times2\) spectral density matrix, maximal coherence is equivalent to
$ \det \mathbf{J}(\omega)=0$, and therefore \(\mathbf{J}(\omega)\) has rank one wherever it is nonzero. Thus, \(\mathbf{J}(\omega)\) defines frequency-resolved bright and dark coupling directions. Nevertheless, these pointwise dark directions need not define a global dark channel: the family \(\{\mathbf{J}(\omega)\}_{\omega}\) may have a trivial common kernel. For the real cross spectra considered here, this occurs when the ratio of the two diagonal entries, i.e. \(J_{22}(\omega)/J_{11}(\omega)\), varies with frequency.  We term this geometric variation \emph{spectral twisting} and
quantify it through the Fubini--Study angle $\Theta(\omega,\omega')$ and the local twist rate $\tau(\omega)$ of the bright projector $\mathbf P_{\mathrm b}(\omega)$.  We determine how twisting modifies dark-state protection and distinguish its effects from those of coupling asymmetry, qubit detuning, thermal absorption, and counter-rotating processes by comparing full and rotating-wave dynamics with suitably constructed untwisted reservoirs.

 As a concrete realization, we consider two qubits coupled through mismatched Drude--Lorentz spectra. We derive analytical expressions for their spectral twisting and calculate the nonperturbative dynamics using HEOM generalized to cross-correlated bath forces. For resonant qubits tuned to the spectral crossing \(\omega_\times\), defined by the condition
\(J_{11}(\omega_\times)=J_{22}(\omega_\times)\), the singlet is locally dark at the transition frequency but remains coupled to off-resonant spectral components. Twisting consequently produces leakage and reduces the persistence of singlet-centered Werner-state concurrence in the parameter regime considered. 
The weak-twisting leakage obeys the fixed-time quadratic scaling $P_{\mathrm{leak}}(t) \propto [\omega_\times\tau(\omega_\times)]^2,
$ before dynamical saturation becomes relevant.

These results provide criteria for dark-state engineering in structured reservoirs and may be relevant to correlated-noise spectroscopy and decoherence-free encodings.

The paper is organized as follows. Section \ref{sec:Model} introduces the common-bath model, the matrix-valued spectral density, and the associated frequency-resolved bright and dark modes. Section \ref{sec:bright_dark_states_common_cutoff} analyzes the globally untwisted couplings and the conditions for exact dark-state protection, while Sec. \ref{sec:spectral_twisting} develops the geometric characterization of spectral twisting. Section \ref{sec:DL_realization} introduces the Drude--Lorentz realization of the spectral density matrix, the results of which are discussed in Sec. \ref{sec:numerical_experiment}, while Sec. \ref{sec:conclusions} summarizes the main findings. Appendix \ref{app:heom_derivation} derives the multichannel HEOM used in the simulations, while Appendix \ref{app:HEOM_validation} derives the secular GKLS generator for correlated channels and an exactly solvable pure-dephasing model that we used to validate the HEOM. Appendix \ref{app:StatRef} discusses the weak-coupling stationary reference and the long-time limit, which provides the weak-coupling stationary-leakage reference used in Sec. \ref{sec:numerical_experiment}.

\section{Two qubits interacting with a common bosonic bath}
\label{sec:Model}

\begin{figure}
    \centering
    \includegraphics[width=\linewidth]{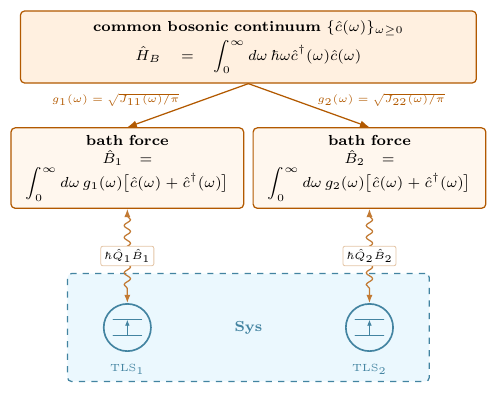}
    {\caption{Two qubits couple to a common bosonic reservoir initially in a thermal state. In the single-continuum construction considered here, both bath-force operators, $\hat B_1$ and $\hat B_2$, couple to the same reservoir mode at each frequency, yielding a rank-one spectral density matrix. The system--reservoir coupling can therefore be decomposed into frequency-resolved bright and dark collective modes.}
\label{fig:rank-one-bath}}
    \label{fig:placeholder}
\end{figure}

We consider two qubits linearly coupled to a common bosonic reservoir through two correlated interaction channels. The system Hilbert space is $\mathcal H_S=\mathcal H_{S_1}\otimes\mathcal H_{S_2}$, and the total Hilbert space is $\mathcal H=\mathcal H_S\otimes\mathcal H_B$, where $\mathcal H_B$ denotes the reservoir Hilbert space. An operator acting locally on qubit $S_1$, for example, is understood as $ \hat O_1\otimes \hat I_{S_2}\otimes \hat I_B$, with analogous conventions for operators acting on the other subsystems. Tensor products with identity operators on the complementary subsystems are left implicit.

\subsection{Common-bath Hamiltonian and spectral density matrix}

The Hamiltonian of the system plus the reservoir, illustrated schematically in Fig. \ref{fig:rank-one-bath}, is
\begin{equation}
\hat H = \hat H_S+\hat H_B+\hat H_I.
\label{eq:Hamiltonian_general}
\end{equation}

The bare system Hamiltonian is
\begin{equation}
\hat H_S = \frac{\hbar\omega_1}{2}\hat\sigma_z^{(1)} + \frac{\hbar\omega_2}{2}\hat\sigma_z^{(2)}
\label{eq:H_sys}
\end{equation}
where $\omega_i$ is the transition frequency of qubit $i=1,2$ and $\hat\sigma_\mu^{(i)}$, with $\mu=x,y,z$ denoting the corresponding Pauli operators. Here, $\ket{0}$ and $\ket{1}$ denote the ground and excited states,
respectively, with $\hat\sigma_z=\ket{1}\!\bra{1}-\ket{0}\!\bra{0}$, $\hat\sigma_+=\ket{1}\!\bra{0}$, $\hat\sigma_-=\ket{0}\!\bra{1}$, and $\hat\sigma_y=i(\hat\sigma_--\hat\sigma_+)$.

The bare reservoir Hamiltonian is
\begin{equation}
\hat H_B = \int_0^\infty d\omega\, \hbar\omega\,\hat c^\dagger(\omega)\hat c(\omega),
\label{eq:rank_one_bath_hamiltonian}
\end{equation}
where the annihilation and creation operators $\hat c(\omega)$ and $\hat c^\dagger(\omega)$ obey the commutation relation $\left[\hat c(\omega),\hat c^\dagger(\omega')\right] = \delta(\omega-\omega')$.

The interaction Hamiltonian $\hat H_I$ is bilinear in the system and reservoir operators,
\begin{equation}
\hat H_I = \hbar \hat Q_1 \hat B_1 + \hbar \hat Q_2 \hat B_2,
\label{eq:H_int}
\end{equation}

where the Hermitian operator $\hat Q_i$ acts locally on qubit $i$, and $\hat B_i$ is the corresponding reservoir-force operator. Both coupling channels are generated by the same bosonic continuum,
\begin{equation}
    \hat B_i=\int_0^\infty d\omega[ g_i(\omega)\hat c({\omega})+ g_i^*(\omega)\hat c^\dagger({\omega})], \qquad i=1,2,
\end{equation}
where $g_i(\omega)$ is the coupling function of channel $i$. We assume that: a) the coupling functions are real and non-negative, $g_i(\omega)\geq0$; b) the initial system--reservoir quantum state is factorized; c) the reservoir is initially in a thermal quantum state at temperature $T$.  
Assumption a) sets the relative cross-spectral phase to zero, leaving only frequency-dependent coupling amplitudes. Assumptions b)--c) are those required by the HEOM construction of Appendix \ref{app:heom_derivation}. 

Since the two forces $\hat B_1$ and $\hat B_2$ involve the same reservoir, their fluctuations are generally cross correlated. We characterize these correlations through
\begin{equation}
C_{ij}(t) = \left\langle \bar B_i(t)\bar B_j(0) \right\rangle_B \qquad i,j=1,2,
\end{equation}
where $\bar B_i(t)$ denotes the reservoir-force operator $\hat B_i$ in the interaction picture generated by $\hat{H}_B$, and $\langle\,\cdot\,\rangle_B=\Tr_B\{\,\cdot\,\hat\rho_B\}$.

Because both coupling channels are generated by the same bosonic continuum, the spectral density matrix 
\begin{equation}
\mathbf J(\omega) =
\begin{pmatrix}
J_{11}(\omega)
&
J_{12}(\omega)
\\[1mm]
J_{21}(\omega)
&
J_{22}(\omega)
\end{pmatrix},
\label{eq:general_spectral_density}
\end{equation}
has entries $ J_{ij}(\omega)=\pi g_i(\omega)g_j(\omega) $. The diagonal elements $J_{11}(\omega)$ and $J_{22}(\omega)$ describe the \textit{local} spectra associated with the two coupling channels, whereas the off-diagonal elements $J_{12}(\omega)$ and $J_{21}(\omega)$ encode their \textit{cross}-spectral correlations. Hence, wherever $\mathbf J(\omega)\neq0$, it has rank one, and under the assumption a), $\mathbf J(\omega)$ is real and symmetric with non-negative entries, $J_{ij}(\omega)\geq0$ and $J_{ij}(\omega)=J_{ji}(\omega)$.  

The correlation function $C_{ij}(t)$ is given by
\begin{multline}
C_{ij}(t) =  \\ \frac{1}{\pi} \int_0^\infty d\omega\, J_{ij}(\omega) \Big[ \bigl(n_\beta(\omega)+1\bigr) e^{-i\omega t} + n_\beta(\omega) e^{i\omega t} \Big],
\label{eq:bath_correlations_spectral_matrix}
\end{multline}
where $n_\beta(\omega) = 1/{(e^{\beta\hbar\omega}-1)}$ is the Bose--Einstein occupation number and $\beta=1/(k_B T)$.

\subsection{Frequency-resolved bright and dark coupling modes}\label{sec:bright_and_dark}
Wherever $\mathbf J(\omega)\neq0$, the rank-one spectral density matrix of Eq. \eqref{eq:general_spectral_density} admits the outer-product factorization $ \mathbf J(\omega) = \mathbf v(\omega)\mathbf v^{\mathrm{T}}(\omega)$,
with $\mathbf{v}(\omega) = \left( \sqrt{J_{11}(\omega)}, \, \sqrt{J_{22}(\omega)} \right)^{\mathrm{T}}$.
For $J_{11}(\omega)+J_{22}(\omega)>0$, the eigenvalues of $\mathbf J(\omega)$ are 
\begin{equation}
J_{\mathrm b}(\omega) = J_{11}(\omega)+J_{22}(\omega), \qquad J_{\mathrm d}(\omega) = 0,
\label{eq:Eigenvalues}
\end{equation}
with corresponding normalized eigenvectors
\begin{subequations}
\begin{align}
\mathbf u_{\mathrm b}(\omega) &= \frac{1}{\sqrt{J_b(\omega)}} ( \sqrt{J_{11}(\omega)}, +\sqrt{J_{22}(\omega)} )^\mathrm{T},
\label{eq:bright_eigenvector}
\\
\mathbf u_{\mathrm d}(\omega) &= \frac{1}{\sqrt{J_b(\omega)}} ( \sqrt{J_{22}(\omega)}, \, -\sqrt{J_{11}(\omega)})^\mathrm{T}.
\label{eq:dark_eigenvector}
\end{align}
\end{subequations}
The labels ``$\mathrm b$'' and ``$\mathrm d$'' denote the bright and dark coupling modes, respectively. These eigenvectors are orthonormal with respect to the ordinary scalar product in $\mathbb R^2$.

Introducing the vector of local coupling operators $ \mathbf{\hat Q} = (\hat Q_1,\hat Q_2 )^{\mathrm{T}}$, the corresponding frequency-resolved collective system operators are
\begin{subequations}
\label{eq:coup_collective_freq_dip}
\begin{align}
\label{eq:coup_collective_freq_dip_a}
\hat Q_{\mathrm b}(\omega) &= \mathbf u_{\mathrm b}^{\mathrm{T}}(\omega)\mathbf{\hat Q}
= \frac{\sqrt{J_{11}(\omega)}\,\hat Q_1 + \sqrt{J_{22}(\omega)}\,\hat Q_2}{ \sqrt{J_b(\omega)}},
\\
\label{eq:coup_collective_freq_dip_b}
\hat Q_{\mathrm d}(\omega) &=\mathbf u_{\mathrm d}^{\mathrm{T}}(\omega)\mathbf{\hat Q} 
=\frac{\sqrt{J_{22}(\omega)}\,\hat Q_1- \sqrt{J_{11}(\omega)}\,\hat Q_2}{\sqrt{J_b(\omega)}}.
\end{align}
\end{subequations}
In terms of them, the interaction Hamiltonian reads:
\begin{equation}
\hat H_I = \hbar \int_0^\infty d\omega\,\sqrt{\frac{J_b(\omega)}{\pi}} \hat Q_{\mathrm b}(\omega) \left[\hat c(\omega) +\hat c^\dagger(\omega) \right].
\label{eq:bright_channel_interaction}
\end{equation}
Hence, only the eigenvector $\mathbf u_{\mathrm b}(\omega)$ contributes to the coupling of the system with the bosonic bath through the collective bright operator $\hat Q_{\mathrm b}(\omega)$. We call $\mathbf u_{\mathrm b}(\omega)$ the \textit{bright mode} of the coupling. The collective dark operator $\hat Q_{\mathrm d}(\omega)$, involving the \textit{dark mode} $\mathbf u_{\mathrm d}(\omega)$, does not appear in the interaction Hamiltonian.

On any connected spectral interval over which $J_{11}(\omega),$  $J_{22}(\omega)>0$, \textit{frequency-independent} bright and dark modes exist if and only if
\begin{equation}
\label{eq:freq_ind_cond}
\frac{J_{11}(\omega)}{J_{22}(\omega)} =
\mathrm{const.}
\end{equation}
throughout that interval.
Equivalently, Eq. \eqref{eq:freq_ind_cond} holds if and only if the family of spectral density matrices possesses a nontrivial common kernel over $I$, $\cap_{\omega\in I}\ker\mathbf J(\omega)\neq\{\mathbf 0\}$.
In this case, the system is related to the bath by an \textit{untwisted collective coupling}.
When this condition is violated, the bright and dark modes rotate (twist) with frequency.  Consequently, although $\mathbf J(\omega)$ remains rank one pointwise, no single frequency-independent transformation diagonalizes the spectral density matrix over the entire interval. We refer to this frequency-dependent rotation of the collective coupling direction as \textit{spectral twisting}.

Henceforth, we consider transverse local couplings,
$\hat Q_i=\hat\sigma_y^{(i)}=i(\hat\sigma_-^{(i)}-\hat\sigma_+^{(i)})$,
with $i=1,2$. Defining the weighted collective lowering operator
\begin{equation}
\hat L_{\rm b}(\omega)
=
\frac{\sqrt{J_{11}(\omega)}\,\hat\sigma_-^{(1)}
+\sqrt{J_{22}(\omega)}\,\hat\sigma_-^{(2)}}
{\sqrt{J_b(\omega)}},
\label{eq:frequency_resolved_collective_lowering}
\end{equation}
we express the \textit{collective bright operator} in
Eq.~\eqref{eq:coup_collective_freq_dip_a} as
\begin{equation}
\hat Q_{\rm b}(\omega)
=
i\left[\hat L_{\rm b}(\omega)-\hat L_{\rm b}^\dagger(\omega)\right].
\label{eq:frequency_resolved_Qb_lowering}
\end{equation}

\section{Dark-state protection with untwisted collective coupling}\label{sec:bright_dark_states_common_cutoff}

We first consider a bosonic reservoir with
\begin{equation}
J_{ii}(\omega) = \eta_i^2 J_0(\omega), \qquad  i=1,2,
\label{eq:proportional_spectral_densities}
\end{equation}
where $\eta_i > 0$ are frequency-independent weights and $J_0(\omega)\geq0$ is a common spectral profile. We impose the normalization
$\eta_1^2+\eta_2^2=1$, which is not restrictive because the overall
normalization can be absorbed into $J_0(\omega)$.

Since $J_{11}(\omega)/J_{22}(\omega)$ is frequency independent, the condition in Eq. \eqref{eq:freq_ind_cond} is satisfied, and the bath is globally untwisted.
The bright eigenvalue is $J_{\rm b}(\omega) = J_0(\omega)$, while
the collective coupling operators are frequency-independent: 
\begin{subequations}
\label{eq:coup_collective_freq_ind}
\begin{align}
\hat Q_{\rm b} &=  \eta_1 \hat Q_1 + \eta_2\hat Q_2,
\\
\hat Q_{\rm d} &=  \eta_2\hat Q_1 - \eta_1\hat Q_2.
\end{align}
\end{subequations}
The interaction Hamiltonian, therefore, reduces to
\begin{equation}
\hat H_I = \hbar \hat Q_{\rm b} \int_0^\infty d\omega\, \sqrt{\frac{J_0(\omega)}{\pi}}\, \left[\hat c(\omega)+\hat c^\dagger(\omega)\right].
\label{eq:global_bright_interaction}
\end{equation}

We introduce the product states $\ket{\phi;\chi}=\ket{\phi}\otimes\ket{\chi}$ where $\ket{\phi}$ belongs to $\mathcal{H}_S$ and $\ket{\chi}$ belongs to $\mathcal{H}_B$. 
A product state $\ket{\phi;\chi}$ is \textit{dark} with respect to the full interaction whenever $\ket{\phi}\in\ker\hat Q_{\rm b}$, i.e.
$\hat Q_{\rm b}\ket{\phi}=0$. In this case,
$\hat H_I\ket{\phi;\chi}=0$ for any reservoir state $\ket{\chi}$, and the
interaction does not couple the system and reservoir. 

Exact \textit{protection} of a particular dark state additionally requires
its free evolution under $\hat H_S$ to remain in the dark subspace
$\mathcal H_D=\ker\hat Q_{\rm b}$. Protection of the entire dark subspace
requires $\hat H_S\mathcal H_D\subseteq\mathcal H_D$.
In particular, a state $\ket{\phi;\chi}$ satisfying
\begin{equation}
\hat Q_{\rm b}\ket{\phi}=0,
\qquad
\hat H_S\ket{\phi}=E_\phi\ket{\phi},
\label{eq:exact_protection_condition}
\end{equation}
is an exactly protected stationary state of the system.

We now specialize to transverse local couplings. The weighted collective lowering operator in Eq. \eqref{eq:frequency_resolved_collective_lowering} becomes
\begin{equation}
\hat L_{\rm b} = \eta_1\hat\sigma_-^{(1)} + \eta_2\hat\sigma_-^{(2)},
\label{eq:weighted_collective_lowering}
\end{equation}
so that Eq. \eqref{eq:frequency_resolved_Qb_lowering} becomes
\begin{equation}
\hat Q_{\rm b} = i \left( \hat L_{\rm b}-\hat L_{\rm b}^\dagger \right).
\label{eq:Qb_lowering_representation}
\end{equation}
\subsubsection{Symmetric coupling and exact protection}
For symmetric couplings, $\eta_1=\eta_2$, the kernel of
$\hat Q_{\rm b}$ is two-dimensional and is spanned by the Bell states
\begin{subequations}
\label{eq:weighted_bright_dark_states}
\begin{align}
\ket{\Psi^-} &= \frac{\ket{10} - \ket{01}}{\sqrt{2}},
\label{eq:weighted_bright_states}
\\
\ket{\Phi^+} &= \frac{\ket{00} +\ket{11}}{\sqrt{2}}.
\label{eq:symmetric_phi_plus_state}
\end{align}
\end{subequations} 
For positive transition frequencies, the state $\ket{\Phi^+}$ is not an eigenstate of $\hat H_S$ and is therefore not protected by the complete Hamiltonian.  The singlet $\ket{\Psi^-}$ is an eigenstate of $\hat H_S$ only for resonant qubits, $\omega_1=\omega_2$. Hence, for symmetric couplings and resonant qubits, $\ket{\Psi^-}$ is protected under the full system--reservoir Hamiltonian. For unequal positive transition frequencies, $\omega_1\neq\omega_2$, no nontrivial stationary protected state exists.

\subsubsection{Asymmetric coupling and lowering-dark states}
For asymmetric couplings, $\eta_1\neq\eta_2$, the kernel of $\hat Q_{\rm b}$ is trivial, $\ker\hat Q_{\rm b}=\{0\}$. The full Hermitian interaction, therefore, possesses no nontrivial dark state. Nevertheless, the state
\begin{equation}
\ket{\rm D}  = \sin\theta\,\ket{10} - \cos\theta\,\ket{01}.
\label{eq:weighted_dark_states}
\end{equation}
belongs to the kernel of the lowering operator $\hat{L}_b$ if
\begin{equation}
\theta = \arctan\sqrt{\frac{J_{22}(\omega)}{J_{11}(\omega)}}  =  \arctan{\frac{\eta_2}{\eta_1}},
\label{eq:untwisted_weighted_angle}
\end{equation}
where $0<\theta<\frac{\pi}{2}$. Because $\eta_1^2+\eta_2^2=1$, it follows that $\cos\theta=\eta_1$ and $\sin\theta=\eta_2$. Consequently,
\begin{equation}
\hat L_{\rm b}\ket D = 0
\qquad \text{and} \qquad
\hat L_{\rm b}^\dagger\ket D = {(\eta_2^2-\eta_1^2)} \ket{11}.
\label{eq:weighted_ladder}
\end{equation}
We call a state satisfying $\hat L_{\rm b}\ket\psi=0$ a \emph{lowering-dark state}. Thus, $\ket D$ is dark to collective emission, but not to the full Hermitian interaction.

Within the rotating-wave approximation, the interaction Hamiltonian is 
\begin{equation}
\hat H_I^{\rm RWA} = i\hbar \int_0^\infty d\omega\, \sqrt{\frac{J_0(\omega)}{\pi}} \left[\hat L_{\rm b}\hat c^\dagger(\omega) - \hat L_{\rm b}^\dagger\hat c(\omega) \right].
\label{eq:bright_interaction_RWA}
\end{equation}
If the reservoir is initially in the vacuum state, the product state $\ket D\otimes\ket{0}_B$ is annihilated by $\hat H_I^{\rm RWA}$: the emission term vanishes because $\hat L_{\rm b}\ket D=0$, while the absorption term vanishes because $\hat c(\omega)\ket{0}_B=0$.

For a vacuum reservoir, exact protection under the total Hamiltonian within the RWA also requires the free evolution of $\ket D$ to remain in $\ker\hat L_{\rm b}$. Directly,
\begin{multline}
\hat H_S\ket D = \hbar \frac{(\omega_1-\omega_2)}{2}\left[ (\eta_2^2-\eta_1^2)\ket D + 2{\eta_1\eta_2}\ket{\rm B} \right]
\label{eq:free_dark_bright_mixing}
\end{multline}
where
$ \ket{\rm B} = {\eta_1\ket{10} + \eta_2\ket{01}}$. Therefore, at zero-temperature and within the RWA, $\ket D$ is exactly protected for resonant qubits, $\omega_1=\omega_2$. For detuned qubits, $\hat H_S$ coherently mixes $\ket D$ and $\ket B$, whereas at finite temperature absorption through $\hat L_{\rm b}^\dagger$ provides an additional leakage channel whenever $\eta_1\neq\eta_2$.

\section{Spectral twisting of collective coupling modes}\label{sec:spectral_twisting}

We now consider the case in which the condition in Eq. \eqref{eq:freq_ind_cond} is not satisfied and the eigendirections of $\mathbf{J}$ become frequency dependent.

On a connected spectral interval $\mathcal I$ where $J_{11}(\omega),J_{22}(\omega)>0$, the frequency-resolved bright and dark coupling modes can be parameterized as
\begin{subequations}
\begin{align}
\mathbf u_{\rm b}(\omega) &=  ( \cos\theta(\omega), +\sin\theta(\omega))^\mathrm{T},
\label{eq:bright_angular_parametrization} \\
\mathbf u_{\rm d}(\omega) &= (\sin\theta(\omega),-\cos\theta(\omega))^\mathrm{T},
\label{eq:orthogonal_dark_eigenvectors}
\end{align}
\end{subequations}
where the mixing angle $0<\theta(\omega)<\frac{\pi}{2}$ is given by:
\begin{equation}
\theta(\omega) = \arctan\sqrt{\frac{J_{22}(\omega)}{J_{11}(\omega)}}.
\label{eq:bright_mixing_angle}
\end{equation}
On the same interval, we define the \textit{local spectral imbalance}
\begin{equation}
\Delta_J(\omega) = \frac{J_{22}(\omega)-J_{11}(\omega)}{J_{11}(\omega)+J_{22}(\omega)},
\label{eq:local_spectral_imbalance}
\end{equation}
where $-1<\Delta_J(\omega)<1$. The mixing angle is then
\begin{equation}
\theta(\omega) = \frac{\pi}{4} + \frac{1}{2}\arcsin\Delta_J(\omega).
\label{eq:general_bright_mixing_angle}
\end{equation}
Thus, the collective direction is frequency independent if and only if $\Delta_J(\omega)$ is constant. Otherwise, the bright and dark directions rotate with frequency, and no single frequency-independent transformation diagonalizes $\mathbf J(\omega)$ throughout $\mathcal I$.

We denote by $\omega_\times$ a positive frequency at which the two local spectral densities cross: $J_{11}(\omega_\times)=J_{22}(\omega_\times)$. At the crossing, $\Delta_J(\omega_\times)=0$ and $\theta^\times=\pi/4$.

\subsection{Frequency-resolved lowering-dark states}
\label{sec:frequency_dependent_dark_direction}

\begin{figure*}
    \centering
    \includegraphics[width=\linewidth]{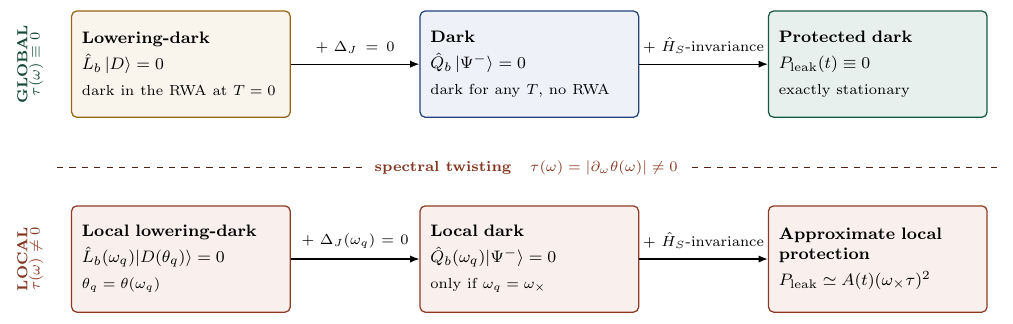}
    \caption{Hierarchy of darkness conditions on the single-excitation manifold $\mathcal H_1=\mathrm{span}\{\ket{10},\ket{01}\}$, with $\ket{D}=\sin\theta\,\ket{10}-\cos\theta\,\ket{01}$ and $\tan\theta=\sqrt{J_{22}/J_{11}}$ for resonant qubits, $\omega_1=\omega_2\equiv \omega_q$. Horizontal arrows add successive conditions, while the dashed line separates untwisted and twisted baths. For an untwisted bath, a lowering-dark state cannot emit but generally satisfies $\hat L_b^\dagger\ket{D}
    =\Delta_J\ket{11}$ where $\Delta_J$ is the spectral imbalance defined by Eq. \eqref{eq:local_spectral_imbalance}; $\Delta_J=0$ makes it fully dark, and invariance under $\hat H_S$ yields exact protection. For a twisted bath, the corresponding conditions hold only locally at $\omega_q$, and protection depends on the spectral window sampled by the dynamics. This nesting is restricted to $\mathcal H_1$.}
    \label{fig:darkness_hierarchy}
\end{figure*}

The state $\ket{D}$ introduced in the previous section is generalized to the frequency-dependent state
\begin{equation}
\ket{D(\omega)} = \ket{D(\theta(\omega))} = \sin\theta (\omega) \,\ket{10} - \cos\theta  (\omega) \,\ket{01},
\label{eq:angle_parametrized_dark_state}
\end{equation}
where $\theta ( \omega )$ is given by Eq. \eqref{eq:bright_mixing_angle}. By construction,
\begin{equation}
\hat L_{\rm b}(\omega)\ket{D(\omega)}=0,
\label{eq:frequency_resolved_lowering_dark_condition}
\end{equation}
so $\ket{D(\omega)}$ is a \textit{frequency-resolved}  {lowering-dark} state. Its behavior under the raising and the complete coupling is instead
\begin{align}
\hat L_{\rm b}^\dagger(\omega)\ket{D(\omega)} &= \Delta_J(\omega)\ket{11},
\\
\hat Q_{\rm b}(\omega)\ket{D(\omega)} &= -i\Delta_J(\omega)\ket{11}.
\label{eq:frequency_resolved_full_dark_condition}
\end{align}
Therefore, $\ket{D(\omega)}$ is generally lowering-dark but not dark with respect to the complete Hermitian coupling.

At a spectral crossing $\omega=\omega_\times$, the frequency-resolved state then reduces
to the singlet,
\begin{equation}
\ket{D(\omega_\times)} = \ket{\Psi^-},
\label{eq:singlet_at_general_crossing}
\end{equation}
which satisfies
\begin{equation}
\hat Q_{\rm b}(\omega_\times)\ket{\Psi^-}=0.
\label{eq:local_full_dark_singlet}
\end{equation}
The singlet is therefore \textit{locally} dark with respect to the complete Hermitian coupling at the crossing frequency.

For two resonant qubits, $\omega_1=\omega_2\equiv \omega_q$, tuned to the crossing, $\omega_q=\omega_\times$, the singlet is also an eigenstate of $\hat H_S$. Hence, system-Hamiltonian invariance is exact, and any loss
of protection originates entirely from the off-resonant spectral variation of the collective coupling direction, since $\hat Q_{\rm b}(\omega)\ket{\Psi^-}\neq0$ for $\omega\neq\omega_\times$. The singlet may nevertheless remain \textit{approximately protected} when the bath frequencies significantly sampled
by the dynamics span a region over which the collective coupling direction varies only weakly.
The hierarchy connecting global lowering darkness, full-interaction darkness, and dynamical protection, together with its local counterpart in a spectrally twisted bath, is summarized in Fig. \ref{fig:darkness_hierarchy}.


\subsection{Geometric measures of spectral twisting}

On each frequency interval $\mathcal I$ over which $\mathbf J(\omega)$ is nonzero, we introduce the normalized projector onto the bright mode,
\begin{equation}
\mathcal{P}_{\rm b}(\omega) = \frac{\mathbf J(\omega)}
{\operatorname{Tr}\mathbf J(\omega)} = \mathbf u_{\rm b}(\omega) \mathbf u_{\rm b}^{\mathrm{T}}(\omega).
\label{eq:bright_projector}
\end{equation}
The relative rotation of the collective coupling direction between two frequencies $\omega$ and $\omega'$ can then be quantified by the \textit{Fubini--Study angle} between the corresponding bright modes
\cite{provost_riemannian_1980,wootters_statistical_1981},
\begin{multline}
\Theta(\omega,\omega') = \arccos\!\left[
\sqrt{\operatorname{Tr} \left\{\mathcal{P}_{\rm b}(\omega) \mathcal{P}_{\rm b}(\omega')\right\}}\right] =\\ \arccos\!\left| \mathbf u_{\rm b}^{\mathrm{T}}(\omega) \mathbf u_{\rm b}(\omega')\right|.
\label{eq:twisting_angle}
\end{multline}
From the angular parametrization introduced by Eq. \eqref{eq:bright_mixing_angle} it follows,
\begin{equation}
\Theta(\omega,\omega') = \left|\theta(\omega)-\theta(\omega')\right|.
\label{eq:twisting_angle_theta}
\end{equation}
If $\mathbf u_{\rm b}(\omega)$ is differentiable, the \textit{local twist rate} at which the bright direction changes with frequency is given by the \textit{Fubini--Study speed} \cite{provost_riemannian_1980,anandan_geometry_1990}
\begin{equation}
\tau(\omega) = \frac{1}{\sqrt{2}} \left\| \frac{\partial \mathcal P_{\rm b}} {\partial\omega}
\right\|_{\mathrm F} = \sqrt{\partial_\omega\mathbf u^{\mathrm{T}}_{\rm b} \partial_\omega\mathbf u_{\rm b} - \left|\mathbf u^{\mathrm{T}}_{\rm b} \partial_\omega\mathbf u_{\rm b} \right|^2},
\label{eq:local_twisting_rate}
\end{equation}
where $\|\cdot\|_{\mathrm F}$ denotes the Frobenius norm. On an interval where $J_{11}(\omega)$ and $J_{22}(\omega)$ are both nonzero, substitution into Eq. \eqref{eq:local_twisting_rate} gives
\begin{equation}
\tau(\omega) = \left|\partial_\omega\theta(\omega)\right|.
\label{eq:twisting_rate_theta_phi}
\end{equation}
Thus, in our case, spectral twisting originates from the frequency-dependent relative spectral weight.

To connect the spectral geometry with the system dynamics, we consider resonant qubits, $\omega_q=\omega_1=\omega_2$, with their transition frequency chosen at a spectral crossing, $\omega_q=\omega_\times$. The frequency-resolved dark state at the crossing $\ket{D(\theta^\times)}$  then coincides with the singlet $\ket{\Psi^-}$. Introducing the angular displacement from the crossing direction,
\begin{equation}
\epsilon(\omega) = \theta(\omega)-\frac{\pi}{4},
\end{equation}
and the symmetric and antisymmetric combinations
\begin{equation}
\hat Q_\pm = \frac{\hat Q_1\pm\hat Q_2}{\sqrt{2}},
\end{equation}
the frequency-resolved bright operator can be written as
\begin{equation}
\hat Q_b(\omega) = \cos\epsilon(\omega)\hat Q_+ - \sin\epsilon(\omega)\hat Q_-.
\label{eq:bright_operator_twisting_expansion}
\end{equation}
For the transverse coupling considered here, $\hat Q_i=\hat\sigma_y^{(i)}$, the singlet satisfies
$\hat Q_+\ket{\Psi^-}=0$, and therefore 
\begin{equation}
\hat Q_{\rm b}(\omega)\ket{\Psi^-}
=
-i\sqrt{2}\sin\epsilon(\omega)\ket{\Phi^+}.
\label{eq:Qb}
\end{equation}
Hence, although $\epsilon(\omega_\times)=0$, spectral components away from the crossing couple to the singlet through the antisymmetric operator $\hat Q_-$. 
Close to the crossing,
\begin{equation}
\left\| \hat Q_{\rm b}(\omega)\ket{\Psi^-} \right\| = \sqrt{2}\, \tau(\omega_\times) \left|\omega-\omega_\times\right| + \mathcal O\!\left[(\omega-\omega_\times)^2\right].
\label{eq:Qb_weak_f}
\end{equation}
Thus, the Fubini--Study speed fixes the leading sensitivity of the singlet's local darkness to spectral twisting. 

For the factorized initial state
$\ket{\Psi^-}\!\bra{\Psi^-}\otimes\hat\rho_B$, with $\hat\rho_B$
thermal, second-order perturbation theory in the system--bath coupling
gives the leakage
\begin{align}
P_{\rm leak}^{(2)}(t)
&=
\frac{1}{\pi}
\int_0^\infty \dd\omega\,
J_{\rm b}(\omega)
\coth\!\left(\frac{\beta\hbar\omega}{2}\right)
\sin^2\epsilon(\omega)
\nonumber\\
&\quad\times
\left[
F_t(\omega-\omega_\times)
+
F_t(\omega+\omega_\times)
\right].
\label{eq:perturbative_leakage_eps}
\end{align}
 in terms of the filter function
\begin{equation}
F_t(\nu)=\left|\int_0^t \dd s\,e^{i\nu s}\right|^2
=\frac{4\sin^2(\nu t/2)}{\nu^2},
\label{eq:filter}
\end{equation}
which describes the near-resonant and, for the argument
$\omega+\omega_\times$, the counter-rotating contribution; the latter
is absent within the RWA.
Because $\epsilon(\omega_\times)=0$, the integrand vanishes
quadratically at the crossing, where $F_t(0)=t^2$. The leakage
therefore probes the off-resonant twist profile over the spectral
range selected by $J_{\rm b}(\omega)$ and the finite-time filters.

For a one-parameter family with fixed crossing frequency
$\omega_\times$ satisfying
\begin{equation}
\epsilon(\omega;\epsilon_0)
=
\epsilon_0 f(\omega)+\mathcal O(\epsilon_0^3),
\qquad
\label{eq:one_parameter_twist}
\end{equation}
with $f(\omega_\times)=0$ and $ \left|
\omega_\times\partial_\omega f(\omega_\times)
\right|=1$, where $f(\omega)$ is independent of $\epsilon_0$, one has
\begin{equation}
\omega_\times\tau(\omega_\times)
=
|\epsilon_0|+\mathcal O(|\epsilon_0|^3).
\end{equation}
Assume furthermore that, at fixed temperature,
\begin{equation}
J_{\rm b}(\omega;\epsilon_0)
=
J_{\rm b}^{(0)}(\omega)+\mathcal O(\epsilon_0^2).
\end{equation}
Expanding  Eq.~\eqref{eq:perturbative_leakage_eps} then gives
\begin{equation}
P_{\rm leak}^{(2)}(t)
=
\mathcal A(t)
\left[\omega_\times\tau(\omega_\times)\right]^2
+
\mathcal O\!\left(
\left[\omega_\times\tau(\omega_\times)\right]^4
\right),
\label{eq:DL_singlet_leakage_delta_scaling}
\end{equation}
where
\begin{align}
\mathcal A(t)
&=
\frac{1}{\pi}
\int_0^\infty \dd\omega\,
J_{\rm b}^{(0)}(\omega)
\coth\!\left(\frac{\beta\hbar\omega}{2}\right)
f^2(\omega)
\nonumber\\
&\quad\times
\left[
F_t(\omega-\omega_\times)
+
F_t(\omega+\omega_\times)
\right].
\label{eq:A_coefficient}
\end{align}
Thus, the local twist rate fixes the quadratic onset, whereas
$\mathcal A(t)$ depends on the full twist profile and its spectral,
thermal, and temporal weighting. The Drude--Lorentz family introduced
in Sec.~\ref{sec:DL_realization} satisfies
Eq.~\eqref{eq:one_parameter_twist} with
$\epsilon_0=\delta_\gamma/2$; for a generic spectral family,
$\tau(\omega_\times)$ alone does not determine the leakage.

\section{Drude--Lorentz realization of spectral twisting}\label{sec:DL_realization}
\begin{figure*}[t]
    \centering
    \includegraphics[width=0.8\linewidth]{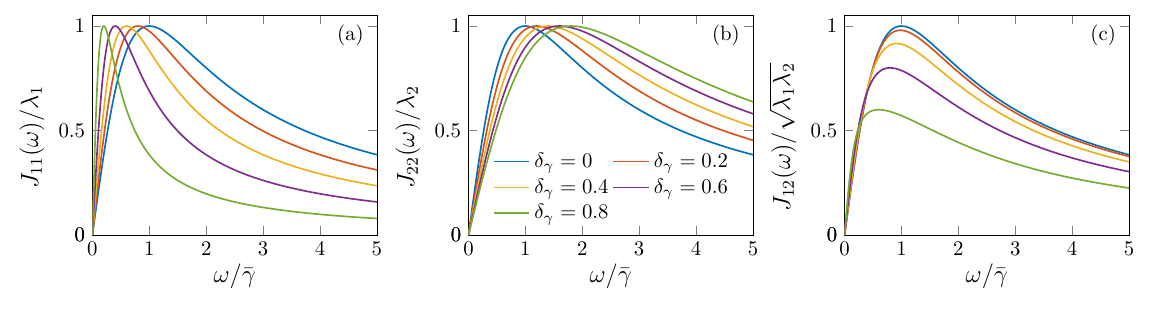}
    \caption{Normalized auto- and cross- spectral densities generated by two local Drude--Lorentz spectra for $\delta_\gamma=0$, $0.2$, $0.4$, $0.6$, and $0.8$.
    Panels (a), (b), and (c) show $J_{11}(\omega)/\lambda_1$,
    $J_{22}(\omega)/\lambda_2$, and $J_{12}(\omega)/\sqrt{\lambda_1\lambda_2}$, respectively, as functions of the normalized frequency $\omega/\bar{\gamma}$, with $\gamma_{1,2}=\bar{\gamma}(1\mp\delta_\gamma)$.}
    
    \label{fig:drude_lorentz_sd}
\end{figure*}

\label{subsec:DL}
As a concrete realization, we consider two Drude--Lorentz spectra,
\begin{equation}
J_{ii}(\omega) = \frac{2\lambda_i\gamma_i\omega}{\omega^2+\gamma_i^2},
\qquad
i=1,2,
\label{eq:general_drude_lorentz_SD}
\end{equation}
where $\lambda_i$ is the reorganization energy and $\gamma_i=\tau_{B_i}^{-1}$ is the bath relaxation rate of channel $i$.
If $\gamma_1=\gamma_2$, the two spectra are proportional and the collective coupling is untwisted, independently of the values of $\lambda_1$ and $\lambda_2$.

To analyze the effects of unequal bath memory times, we set $\lambda_1=\lambda_2$ and introduce the \textit{relaxation rate mismatch}
\begin{equation}
\delta_\gamma = \frac{\gamma_2-\gamma_1}{\gamma_1+\gamma_2}.
\label{eq:cutoff_frequency_mismatch}
\end{equation}
The channels are labeled such that $\gamma_2\geq\gamma_1$. In terms of the mean rate $\bar\gamma=(\gamma_1+\gamma_2)/2$ and $\delta_\gamma$ we have
\begin{equation}
\gamma_1 = \bar{\gamma}(1-\delta_\gamma),
\qquad
\gamma_2 = \bar{\gamma}(1+\delta_\gamma).
\label{eq:cutoffs_in_terms_of_mismatch}
\end{equation}
Figure \ref{fig:drude_lorentz_sd} shows the resulting spectral densities for different values of $\delta_\gamma$.

For $\lambda_1=\lambda_2$ and $\delta_\gamma>0$, the spectra have a unique positive crossing at
\begin{equation}
\omega_\times=\sqrt{\gamma_1\gamma_2}.
\end{equation}
The limit $\delta_\gamma=0$ corresponds to equal bath memory times and identical local spectral densities. Increasing $\delta_\gamma$ produces progressively more asymmetric Drude--Lorentz spectra.
Introducing the dimensionless frequency
\begin{equation}
x=\frac{\omega}{\omega_\times},
\end{equation}
so that the spectral crossing is located at $x=1$, substituting the Drude--Lorentz spectral densities into the expression for the mixing angle in Eq. \eqref{eq:bright_mixing_angle} gives
\begin{equation}
\theta(x;\delta_\gamma)
= \frac{\pi}{4}+\frac{1}{2}\arcsin\left[\delta_\gamma
\frac{x^2-1}{x^2+1}\right].
\label{eq:geometrically_normalized_twisting_angle}
\end{equation}
The corresponding spectral imbalance given by Eq. \eqref{eq:local_spectral_imbalance} is
\begin{equation}
\Delta_J(x)
=
\delta_\gamma
\frac{
x^2-1
}{
x^2+1
}.
\label{eq:drude_lorentz_local_spectral_imbalance}
\end{equation}
Thus, $\delta_\gamma$ directly controls the rotation of the collective coupling direction across the spectrum.

\begin{figure}[t]
    \centering
    \includegraphics[width=\linewidth]{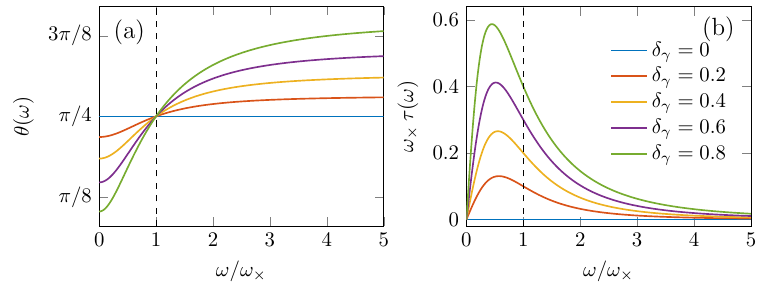}
    \caption{(a) Mixing angle $\theta(\omega)$ for
    $\delta_\gamma=0$, $0.2$, $0.4$, $0.6$, and $0.8$, plotted against
    $\omega/\omega_\times$. For $\delta_\gamma=0$, the collective direction
    is frequency independent, with $\theta=\pi/4$. The vertical dashed line
    marks the spectral crossing $\omega=\omega_\times$, where
    $J_{11}(\omega_\times)=J_{22}(\omega_\times)$ and
    $\theta(\omega_\times)=\pi/4$.
    (b) Corresponding normalized local twisting rate
    $\omega_\times\tau(\omega)$.}
    \label{fig:twisting_plot}
\end{figure}
Figure \ref{fig:twisting_plot}(a) shows $\theta(\omega)$ as a function of  $\omega/\omega_\times$
for different values of $\delta_\gamma$. The normalization by $\omega_\times$ fixes the crossing at $\omega/\omega_\times=1$, allowing the spectral rotation to be compared directly across different $\delta_\gamma$. For $\delta_\gamma=0$, the spectra coincide and $\theta=\pi/4$ at all frequencies; for $\delta_\gamma>0$, the direction rotates monotonically between its low- and high-frequency limits.

Figure \ref{fig:twisting_plot}(b) shows the normalized local twist rate $\omega_\times\tau(\omega)$. For $\delta_\gamma=0$, $\tau(\omega)=0$ at all frequencies. For $\delta_\gamma>0$, the twist rate becomes finite, reaches a maximum below the crossing, and increases as the relaxation-rate mismatch increases. It decreases toward zero in the high-frequency limit.

At the crossing,
\begin{equation}
\tau \left(\omega_\times\right) = \left| \left. \partial_\omega \theta \right|_{\omega=\omega_\times} \right| = \frac{\delta_\gamma}{2\sqrt{\gamma_1\gamma_2}}.
\label{eq:twisting_rate_crossing}
\end{equation}
Thus, $\delta_\gamma$ controls both the angular rotation across the spectrum and the local twisting rate at the crossing.
Several dynamical regimes can be distinguished according to the position of the qubit transition frequency relative to the crossing.

\subsection{Resonant qubits tuned to the spectral crossing}

We first consider two resonant qubits tuned to the crossing, $\omega_q=\omega_\times$.
For $\delta_\gamma=0$ the singlet is dark at every frequency. For $\delta_\gamma>0$, it remains dark only at the crossing:  no frequency-independent lowering-dark state exists over the full spectrum. If the dynamics samples only a sufficiently narrow frequency window around $\omega_\times$, the singlet may nevertheless remain approximately long-lived. 

Starting from  $ \hat\rho_S(0) = \ket{D(\omega_\times)}\bra{D(\omega_\times)}  =\ket{\Psi^-}\bra{\Psi^-}$, we
quantify the loss of protection by
\begin{equation}
P_{\rm leak}(t) = 1- \bra{D(\theta^\times)} \hat\rho_S(t) \ket{D(\theta^\times)}.
\label{eq:angle_resolved_leakage}
\end{equation}
The angular displacement is exactly
\begin{equation}
\epsilon(x;\delta_\gamma) = \frac{1}{2} \arcsin\!\left[\delta_\gamma \frac{x^2-1}{x^2+1} \right].
\label{eq:DL_exact_epsilon}
\end{equation}
In the weak-twisting regime,
\begin{equation}
\epsilon(x) = \frac{\delta_\gamma}{2} \frac{x^2-1}{x^2+1} + \mathcal O\, (\delta_\gamma^3).
\label{eq:DL_epsilon_tau}
\end{equation}
Hence, Eq. \eqref{eq:Qb} gives
\begin{equation}
\hat Q_b(\omega)\ket{\Psi^-} \simeq - \frac{\delta_\gamma}{2} \frac{x^2-1}{x^2+1} \hat Q_-\ket{\Psi^-} +
\mathcal O\, (\delta_\gamma^3).
\label{eq:DL_dark_breaking}
\end{equation}
Under $\delta_\gamma\mapsto-\delta_\gamma$, the two bath channels are exchanged. For identical qubits and a singlet initial state, this exchange leaves the survival leakage invariant. Consequently, its perturbative expansion contains only even powers of $\delta_\gamma$. Since the singlet is exactly protected at $\delta_\gamma=0$, the leading contribution to $P_{\rm leak}$ is given by Eq. \eqref{eq:DL_singlet_leakage_delta_scaling} where the coefficient $\mathcal A(t)$, given by \eqref{eq:A_coefficient}, contains the spectral weighting of the off-resonant modes and the effects of temperature, bath memory, and counter-rotating processes. 

\subsection{High-frequency off-crossing limit}

\label{sec:High_freq}

We next consider resonant qubits whose common transition frequency $\omega_q$ lies well above the spectral crossing, $x_q={\omega_q} / {\omega_\times}\gg1.$
In this regime, as shown in
Fig.~\ref{fig:twisting_plot}(a), the mixing angle approaches its
high-frequency plateau, while panel (b) shows that the local twist rate
$\tau(\omega)$ tends to zero. The collective coupling is therefore \textit{approximately} untwisted over an extended spectral region around $\omega_q$, although it remains globally twisted. This scenario is therefore very similar to the untwisted asymmetric-coupling case discussed in Sec. \ref{sec:bright_dark_states_common_cutoff}, with  $\eta_1^2 = J_{11} (\omega_q) / (J_{11} (\omega_q) + J_{22} (\omega_q)) $ and $\eta_2^2 = J_{22} (\omega_q)/ (J_{11} (\omega_q) + J_{22} (\omega_q))$. 
The locally lowering-dark state at the qubit frequency is obtained from
Eq.~\eqref{eq:weighted_dark_states} by evaluating the angle in
Eq.~\eqref{eq:untwisted_weighted_angle} at $\omega_q$, namely
$\theta_q\equiv\theta(\omega_q)$. For the Drude--Lorentz spectra considered
here, this gives
\begin{equation}
\theta_q=\frac{\pi}{4}+\frac{1}{2} \arcsin \left[\delta_\gamma \frac{x_q^2-1}{x_q^2+1}\right].
\end{equation}
Hence, in the high-frequency limit,
\begin{equation}
    \theta_{\infty}=\frac{\pi}{4}+\frac{1}{2} \arcsin \delta_\gamma.
    \label{eq:AngleHighFreq}
\end{equation}
For large but finite $x_q$, the residual spectral variation produces
corrections to this effective untwisted description.

\section{Nonperturbative two-qubit dynamics}\label{sec:numerical_experiment}
We investigate numerically the model of two qubits coupled to a common bosonic reservoir introduced in Sec. \ref{sec:Model}, with the system Hamiltonian given by Eq. \eqref{eq:H_sys}. We first consider two resonant qubits, $\omega_1=\omega_2\equiv\omega_q$; then we analyze the effects of qubit-frequency detuning.

The system couples to the common environment through $\hat Q_1=\hat\sigma_y^{(1)}$ and $\hat Q_2=\hat\sigma_y^{(2)}$. The local spectra $J_{11}(\omega)$ and $J_{22}(\omega)$ are the Drude--Lorentz functions of Eq. \eqref{eq:general_drude_lorentz_SD}, characterized by reorganization energies $\lambda_1,\lambda_2$ and relaxation rates $\gamma_1,\gamma_2$. We fix the dimensionless temperature to ${k_BT}/({\hbar\omega_q})=0.2$. Unless explicitly identified as RWA results, all simulations use the complete Hermitian system–reservoir interaction.

For the pure-state calculations, we consider the angle-parameterized single-excitation initial state
\begin{equation}
\ket{\psi(\alpha)} = \sin\alpha\,\ket{10} - \cos\alpha\,\ket{01}.
\label{eq:angle_parametrized_dark_state_f}
\end{equation}
Let $\hat\rho_\alpha(t)$ denote the corresponding reduced density operator. We introduce the survival leakage

\begin{equation} 
P_{\rm leak}(\alpha,t) = 1- \bra{\psi(\alpha)} \hat {\rho}_\alpha(t) \ket{\psi(\alpha)}. \label{eq:angle_resolved_leakage_phi} 
\end{equation} 

To identify the initial state that remains optimally protected over a finite observation window \([0,t_f]\), we consider the cumulative time-averaged leakage
\begin{equation}
\overline{P}_{\rm leak}(\alpha;t) = \frac{1}{t} \int_0^{t} P_{\rm leak}(\alpha,t')\,dt',
\label{eq:cumulative_leakage}
\end{equation}
and define the corresponding cumulative optimal angle 
\begin{equation}
\alpha_{\rm opt}(t) = \arg\min_{\alpha} \overline{P}_{\rm leak}(\alpha;t).
\label{eq:cumulative_optimal_angle}
\end{equation}
We calculate the reduced dynamics using an in-house implementation of the HEOM for cross-correlated coupling channels, derived in
Appendix \ref{app:heom_derivation}. In Appendix \ref{app:HEOM_validation}, we discuss the validation of the HEOM code. Frequencies are expressed in units of $\omega_q$, and times in units of $\omega_q^{-1}$.

For $\gamma_1\neq\gamma_2$, the exact rank-one cross spectrum $ J_{12}(\omega)=\sqrt{J_{11}(\omega)J_{22}(\omega)} $ is not itself of Drude-Lorentz form. We approximate it by three Drude components, with the fit constrained to preserve the positive semidefiniteness of $\mathbf J(\omega)$ over the relevant frequency range \cite{garraway1997pseudomode}. Its deviation from exact rank one is quantified by
\begin{equation}
\epsilon_{\rm rank}(\omega) = \frac{\left|J_{11}(\omega)J_{22}(\omega)-|J_{12}^{\rm fit}(\omega)|^2\right|}{J_{11}(\omega)J_{22}(\omega)},
\end{equation}
for which $\max_\omega\epsilon_{\rm rank}(\omega)<10^{-6}$ over the whole frequency range relevant to the simulations. Thus, over this range, the fitted matrix approximates the pointwise rank-one structure to high accuracy and does not introduce an appreciable second spectral channel.
At finite temperature, each local and cross correlation function is represented using a Pad\'e expansion \cite{hu2010pade,hu_pade_2011} of order $N_{\rm P}=11$. Coincident decay rates are combined, the residual high-frequency contribution is included through a terminator correction, and the hierarchy is truncated at depth $N_{\rm C}=3$, as defined in Sec. \ref{sec:ADOsHEOM}.
Numerical convergence was verified in the most demanding parameter regimes, including those with the largest spectral asymmetries, by increasing the expansion order to \(N_{\rm P}=12\) and the hierarchy depth to \(N_{\rm C}=4\). The resulting dynamics showed no appreciable deviations, confirming convergence with respect to both truncation parameters.

For the RWA calculations, we retain the same bath temperature and target spectral density matrix as in the full-interaction simulations but adapt the
correlation decomposition to the interaction Hamiltonian. Whereas the complete Hermitian coupling involves the full bath correlation function,
including both emission and absorption contributions, the RWA couples $\hat{\sigma}_{-}^{(i)}$ and $\hat{\sigma}_{+}^{(i)}$ separately to the
corresponding emission and absorption components, weighted by $n_\beta(\omega)+1$ and $n_\beta(\omega)$, respectively. These correlation
functions are fitted separately using a common set of $12$ positive exponential decay rates.

We first analyze coupling-strength asymmetry in the globally untwisted regime, $\gamma_1=\gamma_2$, in Sec. \ref{subsec:untwisted-coupling-asymmetry}. We then consider the twisted regime obtained for $\lambda_1=\lambda_2$, first for pure initial states in Sec. \ref{subsec:ResLeakage} and then for Werner-type mixed states in Sec. \ref{subsec:werner_states}. In Sec. \ref{subsec:detun}, we analyze the consequences of the qubit-frequency detuning.


\subsection{Leakage from untwisted coupling asymmetry}
\label{subsec:untwisted-coupling-asymmetry}
\begin{figure*}
  \centering
    \includegraphics[width=\linewidth]{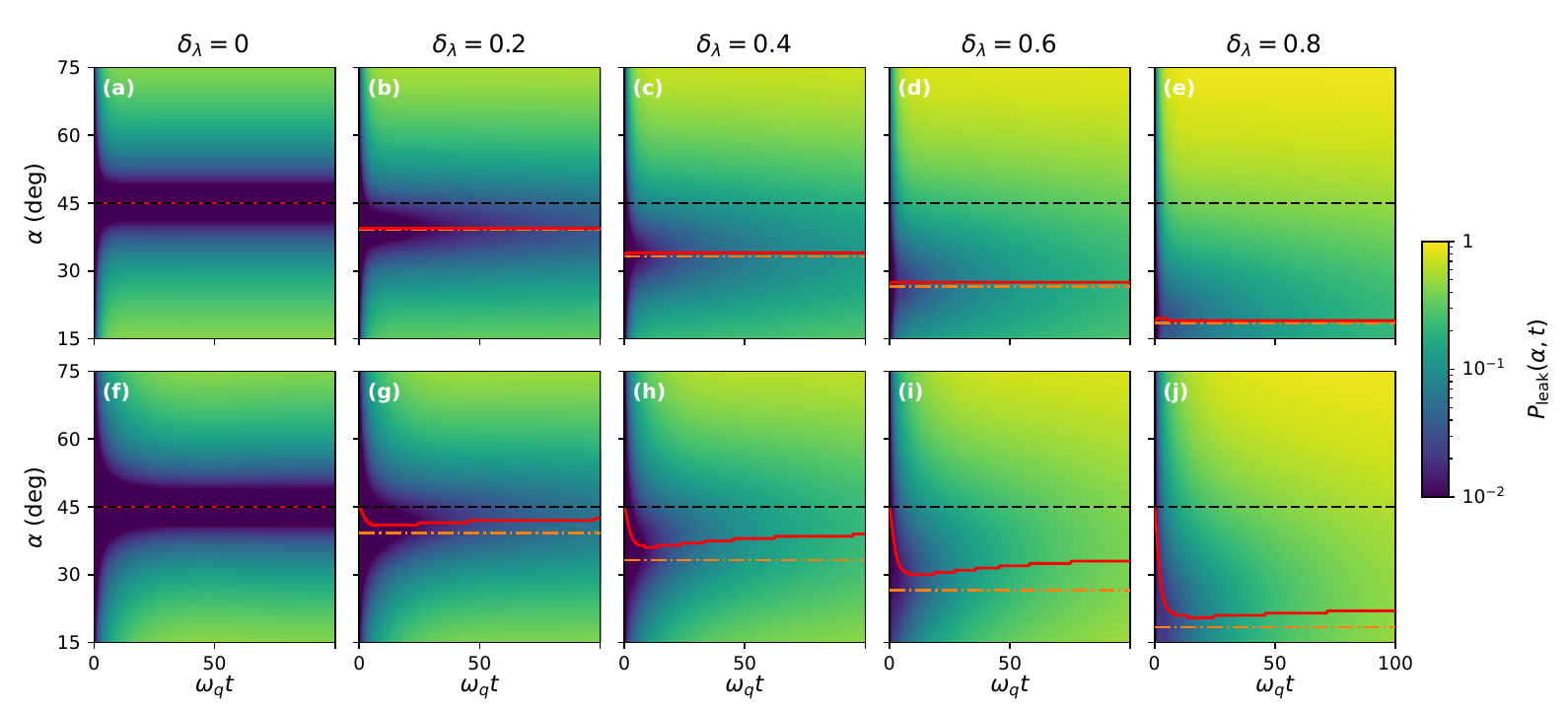}
    \caption{Time-resolved leakage probability $P_{\rm leak}(\alpha,t)$ for the single-excitation initial state
    $\ket{\psi(\alpha)} =\sin\alpha\,\ket{10}-\cos\alpha\,\ket{01}$ in the globally untwisted regime. From left to right, $\delta_\lambda=0$, $0.2$, $0.4$, $0.6$, and $0.8$. Panels (a)--(e) show the RWA dynamics, while panels (f)--(j) show the full Hermitian dynamics. The logarithmic color scale spans $10^{-2}\leq P_{\rm leak}\leq1$; smaller values are clipped to its lowest level. The red solid curve denotes the cumulative optimal angle
    $\alpha_{\rm opt}(t)$, the black dashed line marks the singlet direction $\alpha=\pi/4$, and the orange dash--dotted line marks the frequency-independent lowering-dark angle $\alpha_D=\arctan\sqrt{(1-\delta_\lambda)/(1+\delta_\lambda)}$. In all calculations, $\gamma_1=\gamma_2=\omega_q$, $\Lambda=0.05\,\omega_q$, and $k_BT/(\hbar\omega_q)=0.2$.}
    \label{fig:optimal_leakage_heatmap_untwisted}
\end{figure*}

\begin{figure*}[t]
    \centering
    \includegraphics[width=0.8\linewidth]{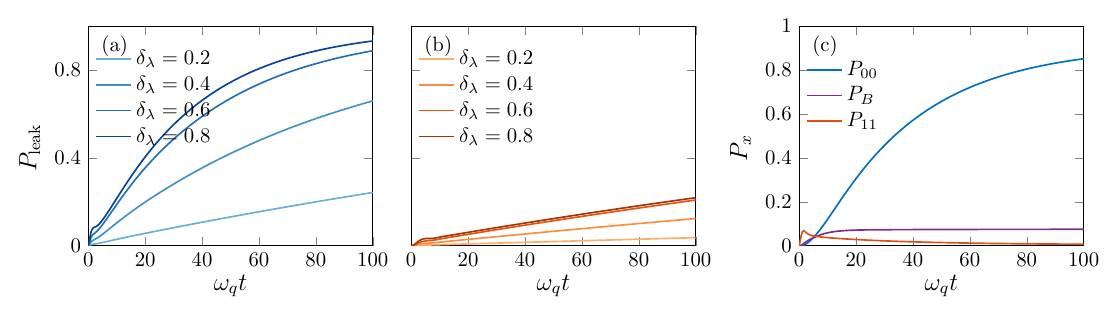}
    \caption{Leakage probability from the lowering-dark initial state $\ket{D}$ for an untwisted common bath with unequal coupling strengths. Panel (a) shows the full Hermitian interaction, while panel (b) shows the rotating-wave result. Panel (c) shows the population dynamics outside the lowering-dark state $\ket D$ for the full Hermitian interaction at $\delta_\lambda=0.8$. The curves show the populations of the ground state, $P_{00}$, the doubly excited state, $P_{11}$, and the weighted bright state, $P_B$, which satisfy the decomposition $P_{\rm leak}(t) = P_{00}(t)+P_B(t)+P_{11}(t)$. In all calculations, $\gamma_1=\gamma_2=\omega_q=1$, $\Lambda=0.2\omega_q$, and $k_BT/\hbar\omega_q=0.2$.}
    \label{fig:untwisted-coupling-asymmetry}
\end{figure*}

We isolate coupling-strength asymmetry in the globally untwisted regime, assuming $\gamma_1=\gamma_2$ and varying
\begin{equation}
\delta_\lambda
=
\frac{\lambda_1-\lambda_2}{\lambda_1+\lambda_2},
\qquad
\lambda_{1,2}
=
\frac{\Lambda}{2}(1\pm\delta_\lambda),
\label{eq:coupling_strength_mismatch}
\end{equation}
at fixed $\Lambda=\lambda_1+\lambda_2$. For $\omega_q=\gamma_1=\gamma_2$, this also fixes the resonant bright spectral weight,
\begin{equation}
J_{\rm b}(\omega_q) = \operatorname{Tr}\mathbf J(\omega_q) = \Lambda.
\end{equation}
Because the two local spectra remain proportional at all frequencies, the bath is globally untwisted and hence
$\tau(\omega)=0$ for every value of $\delta_\lambda$. 
In particular, the lowering-dark state within the parametrization of Eq. \eqref{eq:angle_parametrized_dark_state_f} is obtained for
\begin{equation}
\alpha_D = \arctan\sqrt{\frac{\lambda_2}{\lambda_1}} = \arctan\sqrt{\frac{1-\delta_\lambda}{1+\delta_\lambda}}.
\label{eq:alpha_D_lowering}
\end{equation}

For resonant qubits within the RWA, $\ket D\otimes\ket{0}_B$
is exactly protected. At finite temperature, however, thermal
absorption can populate $\ket{11}$, since
\begin{equation}
\bra D\hat L_{\rm b}\hat L_{\rm b}^\dagger\ket D
=\delta_\lambda^2.
\label{eq:matrix_element_D}
\end{equation}
A weak-coupling resonant-rate estimate therefore gives
\begin{equation}
\Gamma_{\rm leak}^{\rm th}
\propto n_\beta(\omega_q)J_{\rm b}(\omega_q)\delta_\lambda^2
=n_\beta(\omega_q)\Lambda\delta_\lambda^2.
\end{equation}
The full Hermitian interaction also permits leakage from
a vacuum reservoir. Indeed, Eq.~\eqref{eq:Qb_lowering_representation}
gives
\begin{equation}
\hat Q_{\rm b}\ket D=i\delta_\lambda\ket{11},
\label{eq:full_interaction_asymmetry_action}
\end{equation}
so it couples $\ket D\otimes\ket{0}_B$
to states with both a doubly excited qubit pair and a bath
excitation.

Figure~\ref{fig:optimal_leakage_heatmap_untwisted} shows how
these mechanisms affect the angle-resolved dynamics. Panels (a)--(e) show the RWA results, while panels (f)--(j) show the results for the full Hermitian interaction.  For symmetric couplings, i.e. $\delta_\lambda=0$, the singlet is exactly protected in both models. As the coupling asymmetry increases, the low-leakage region shifts toward smaller angles, following the displacement of the frequency-independent lowering-dark direction $\alpha_D$. The cumulative optimal angle $\alpha_{\rm opt}(t)$ exhibits the same trend.  Within the RWA,
$\alpha_{\rm opt}(t)$ remains close to $\alpha_D$, with small
but finite thermal leakage.

Under the full interaction, the optimum generally departs from $\alpha_D$ and the minimum leakage increases with asymmetry due to counter-rotating processes.

Figure \ref{fig:untwisted-coupling-asymmetry} examines in greater detail the differences between the full and RWA dynamics. In order to explore the long-time behavior, we increased the coupling to $\Lambda = 0.2 \omega_q$. In the full model, the leakage at $\omega_q t=100$ increases from approximately $0.242$ at $\delta_\lambda=0.2$ to $0.932$ at $\delta_\lambda=0.8$. The corresponding RWA values are approximately $0.036$ and $0.218$. In the short-time regime, $\omega_q t\lesssim1$, the dependence on the coupling asymmetry is quadratic. For example, at $\omega_q t=1$, the numerical results scale approximately as $P_{\rm leak}^{\rm full}\simeq0.098\,\delta_\lambda^2$ and $P_{\rm leak}^{\rm RWA}\simeq0.0069\,\delta_\lambda^2$.
At longer times, higher-order terms in $\delta_\lambda$ become significant, and the leakage no longer follows the leading quadratic approximation. Figure \ref{fig:untwisted-coupling-asymmetry}(c) shows that most of the population leaving $\ket D$ is subsequently transferred to the ground state. These results demonstrate that a frequency-independent bright direction is not sufficient for exact protection. In the present resonant setting, the candidate state must also be dark with respect to the complete interaction, as established in Sec. \ref{sec:bright_dark_states_common_cutoff}.

\subsection{Leakage induced by spectral twisting}\label{subsec:ResLeakage}

\begin{figure*}
  \centering
    \includegraphics[width=\linewidth]{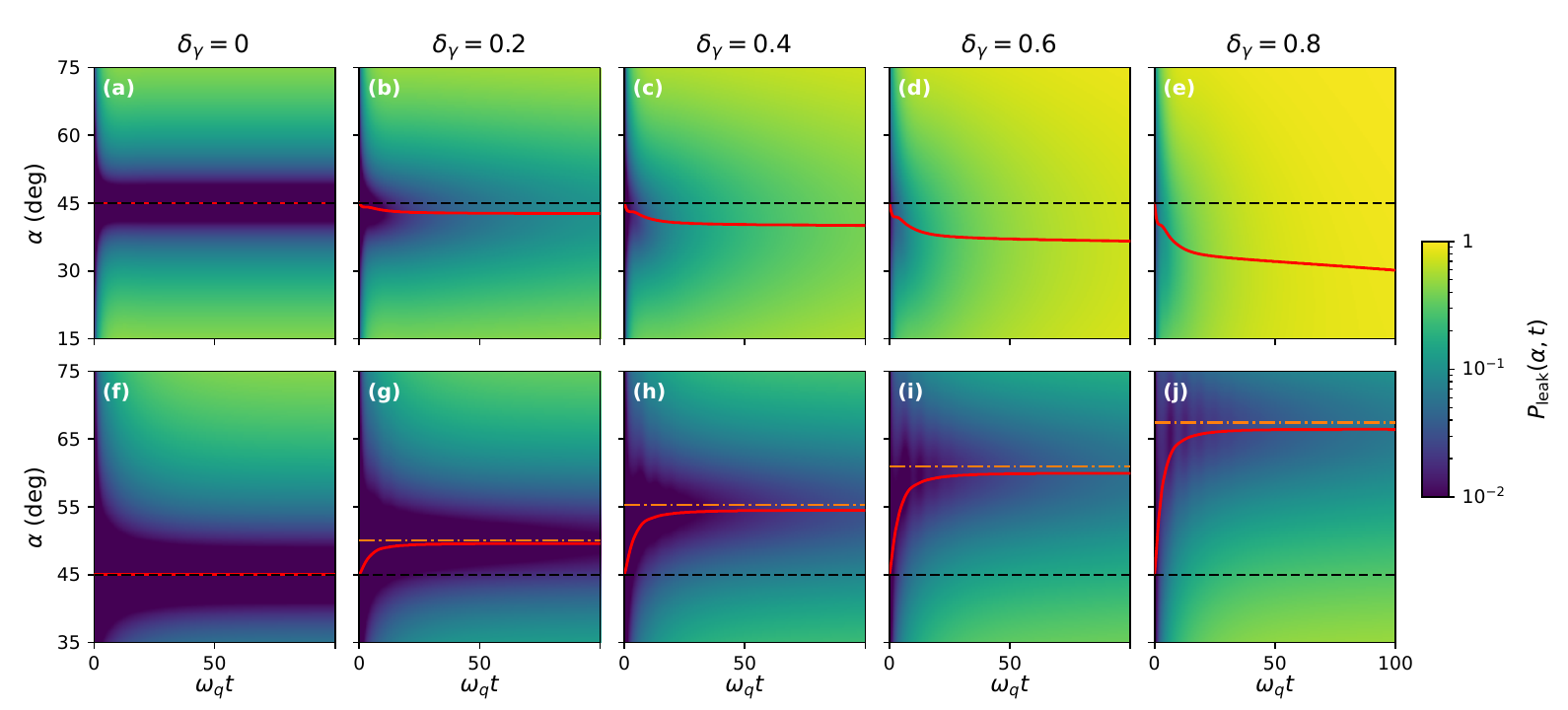}
    \caption{Time-resolved leakage probability $P_{\rm leak}(\alpha,t)$ for the single-excitation initial state $\ket{\psi(\alpha)}$, for $\delta_\gamma=0$, $0.2$, $0.4$, $0.6$, and $0.8$.
    Panels (a)--(e) correspond to the spectral-crossing configuration $\omega_q=\omega_\times$, while panels (f)--(j) correspond to the off-crossing configuration $\omega_q=4\omega_\times$. The logarithmic color scale spans $10^{-2}\leq P_{\rm leak}\leq1$; smaller values are clipped to its lowest level. The red solid curve denotes the cumulative optimal angle $\alpha_{\rm opt}(t)$, the black dashed line marks the singlet direction $\alpha=\pi/4$, and the orange dash--dotted line marks the frequency-resolved lowering-dark angle $\alpha_D(\omega_q)$. In panels (a)--(e) the orange dash-dotted line is omitted since $\alpha_D(\omega_q) = \pi/4$ coincides with the singlet direction. In all calculations,
    $J_{\rm b}(\omega_q)=0.2\,\omega_q$ and $k_BT/(\hbar\omega_q)=0.2$.}
    \label{fig:optimal_leakage_heatmap}
\end{figure*}

We now investigate how the twisting induced by the relaxation-rate mismatch $\delta_\gamma$ affects the  leakage. The lowering-dark angle within the parametrization of Eq. \eqref{eq:angle_parametrized_dark_state_f} at the qubit transition frequency is
\begin{equation}
\alpha_{\rm D}(\omega_q) = \theta_{}(x_q;\delta_\gamma).
\label{eq:dark_angle_at_qubit_result}
\end{equation}
where $\theta$ is given by \eqref{eq:geometrically_normalized_twisting_angle}. Throughout this subsection, the crossing frequency $\omega_\times = \sqrt{\gamma_1\gamma_2}$ is held fixed while $\delta_\gamma$ is varied. We also assume identical reorganization energies, i.e. $\lambda_1 = \lambda_2 = \lambda$, and adjust their common value for each $\delta_\gamma$ so that the resonant bright spectral weight $J_{\rm b}(\omega_q)$ remains fixed to $0.2 \omega_q$ for each value of $\delta_\gamma$,
\begin{equation}
\lambda(\delta_\gamma) = \frac{J_{\rm b} (\omega_q)}{2\omega_q\left[\dfrac{\gamma_1}{\omega_q^2+\gamma_1^2}+\dfrac{\gamma_2}{\omega_q^2+\gamma_2^2}\right]}.
\end{equation}

\subsubsection{Qubits at the spectral crossing}

We first set $\omega_q=\omega_\times$, as in
Figs.~\ref{fig:optimal_leakage_heatmap}(a)--(e). At the crossing,
\begin{equation}
\alpha_{\mathrm D}(\omega_q)
=
\theta(\omega_\times)
=
\frac{\pi}{4},
\qquad
\omega_\times\tau(\omega_\times)
=
\frac{\delta_\gamma}{2}.
\end{equation}
Although the singlet is therefore locally dark for every
$\delta_\gamma$, the HEOM dynamics shows that the cumulative optimal angle
moves progressively below $\pi/4$.  Since the frequency-resolved dark angle
satisfies $\theta(\omega)<\pi/4$ for $\omega<\omega_\times$, this shift
reflects the influence of off-resonant modes below the crossing. The
optimal initial state is therefore determined by the frequency-weighted
coupling over the range of bath modes sampled during the evolution, not
by the resonant spectral matrix $\mathbf J(\omega_q)$ alone.

\begin{figure*}
    \centering
    \includegraphics[width=0.75\linewidth]{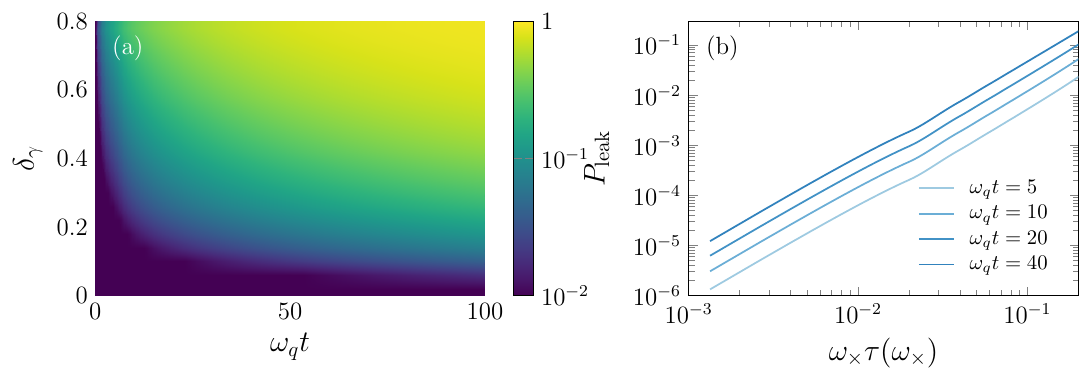}
    \caption{Singlet leakage at the spectral crossing $\omega_q=\omega_\times$. (a) Time-resolved leakage probability $P_{\mathrm{leak}}(t)$ as a function of $\omega_q t$ and the bath relaxation-rate mismatch $\delta_\gamma$. The logarithmic color scale spans $10^{-2}\leq P_{\rm leak}\leq1$. (b) Log--log representation of the fixed-time cuts of panel (a) at $\omega_q t=5$, $10$, $20$, and $40$, plotted as functions of the local twisting parameter $\omega_\times\tau(\omega_\times)=\delta_\gamma/2$ over the weak-twisting interval $0<\omega_\times\tau(\omega_\times)\leq0.2$. The approximately linear behavior in the log--log representation highlights the quadratic weak-twisting scaling of the leakage probability. In all calculations, $J_b(\omega_q)=0.2 \omega_q$ and $k_B T/(\hbar\omega_q)=0.2$.
}
    \label{fig:FigureTwist}
\end{figure*}

Figure~\ref{fig:FigureTwist}(a) shows that the singlet remains protected for
$\delta_\gamma=0$, whereas any finite twisting produces leakage that grows
with both time and $\delta_\gamma$. Because $\mathbf J(\omega_q)$ is fixed throughout the sweep, this leakage is enabled by the coupling of the singlet to the off-resonant collective directions generated by spectral twisting. Its magnitude is additionally controlled by the off-resonant spectral weight, bath memory, temperature, and counter-rotating processes.
The fixed-time cuts in Fig.~\ref{fig:FigureTwist}(b) satisfy
\begin{equation}
P_{\mathrm{leak}}(t)
\propto
\bigl[\omega_\times\tau(\omega_\times)\bigr]^2
\end{equation}
for $\omega_\times\tau(\omega_\times)\lesssim0.2$, in agreement with
Eq.~\eqref{eq:DL_singlet_leakage_delta_scaling}. Deviations at larger twisting
reflect higher-order contributions and the onset of dynamical saturation.

This fixed-time scaling does not describe the stationary limit. As discussed
in Appendix~\ref{app:StatRef}, for every fixed $\delta_\gamma>0$ a thermalizing
dynamics approaches, in the weak-coupling limit,
\begin{equation}
P_{\mathrm{leak}}^{(\beta,0)}
=
1-\frac{1}{4}
\operatorname{sech}^{2}\!\left(\frac{\beta\hbar\omega_q}{2}\right),
\end{equation}
which equals $0.9934$ for $k_BT/(\hbar\omega_q)=0.2$. The limit
$\delta_\gamma\to0$ is therefore singular: the singlet is exactly protected
at $\delta_\gamma=0$, while any finite twisting opens a relaxation channel.
If its slowest rate scales as
$\Gamma(\delta_\gamma)\propto\delta_\gamma^2$, the quadratic fixed-time law
holds for $\Gamma(\delta_\gamma)t\ll1$, whereas at longer times the leakage
saturates toward its stationary value.
\subsubsection{Qubits in the high-frequency off-crossing limit}

We next consider $\omega_q=4\omega_\times$, corresponding to $x_q=4$ and Figs. \ref{fig:optimal_leakage_heatmap}(f)--(j). This places the qubit transition on the high-frequency branch of the spectral rotation, where $\theta(x;\delta_\gamma)$ approaches its asymptotic plateau as shown in Fig. \ref{fig:twisting_plot}.  As shown in Sec. \ref{sec:High_freq}, the reservoir is therefore 
approximately untwisted, although it remains globally twisted, and its
dynamics should resemble that of an untwisted reservoir with asymmetric
couplings.

The corresponding frequency-resolved lowering-dark angle is close to the high-frequency limit given by Eq. \eqref{eq:AngleHighFreq}. Accordingly, the low-leakage region shifts above the singlet angle as
\(\delta_\gamma\) increases. For \(\delta_\gamma=0\), the collective
direction is frequency independent and the singlet remains exactly
protected.

Table~\ref{tab:numerical_optimal_angles} shows that the cumulative
optimum at \(\omega_q t_f=100\) closely follows
\(\alpha_{\rm D}(\omega_q)\), confirming the locally untwisted
description of Sec.~V\,B. The small residual displacement and finite
leakage arise because the dynamics samples a finite spectral window
over which the bright direction is not exactly constant.
Counter-rotating processes further enhance this sensitivity to
off-resonant frequencies.

\begin{figure*}
  \centering
  \includegraphics[width=0.75\linewidth]{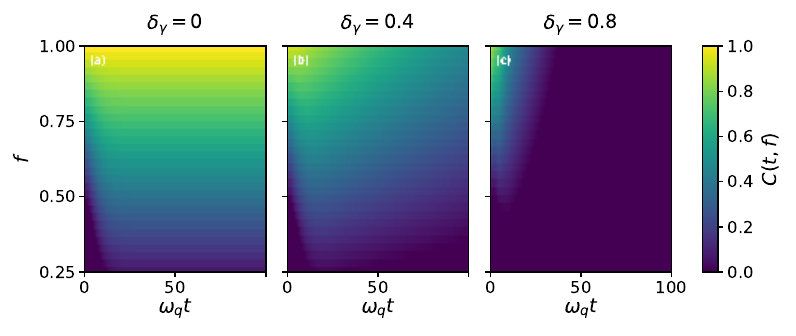}
  \caption{Concurrence $C(t,f)$ of singlet-centered Werner states for $\delta_\gamma=0$, $0.4$, and $0.8$, with $\omega_q=\omega_\times$, $J_b(\omega_q)=0.2\,\omega_q$, and $k_BT/(\hbar\omega_q)=0.2$.}
  \label{fig:werner_concurrence}
\end{figure*}

\begin{table}[t]
\centering
\caption{Optimized angle $\alpha_{\rm opt}$, local lowering-dark angle $\alpha_D(\omega_q)$ and leakage $P_\mathrm{leak}(\alpha_{\rm opt}, \omega_q t_f)$ for different values of the bath relaxation-rate mismatch \(\delta_\gamma\) in the case of $\omega_q = 4\,\omega_\times$.}
\label{tab:numerical_optimal_angles}
\begin{tabular}{c|ccccc}
\hline
\(\delta_\gamma\)
&
\(0\)
&
\(0.2\)
&
\(0.4\)
&
\(0.6\)
&
\(0.8\)
\\
\hline
\(\alpha_{\rm opt}\)
&
\(45.00^\circ\)
&
\(49.60^\circ\)
&
\(54.50^\circ\)
&
\(59.95^\circ\)
&
\(66.40^\circ\)
\\
\(\alpha_D(\omega_q)\)
&
\(45.00^\circ\)
&
\(50.08^\circ\)
&
\(55.33^\circ\)
&
\(60.98^\circ\)
&
\(67.45^\circ\)
\\
\(P_{\rm leak}(\alpha_{\rm opt},\omega_qt_f)\)
&
\(0\)
&
\(9\times10^{-3}\)
&
\(3\times10^{-2}\)
&
\(5\times10^{-2}\)
&
\(6\times10^{-2}\)
\\
\hline
\end{tabular}
\end{table}

\subsection{Werner-state entanglement under spectral twisting}
\label{subsec:werner_states}

\begin{figure*}[t]
    \centering
    \includegraphics[width=\linewidth]{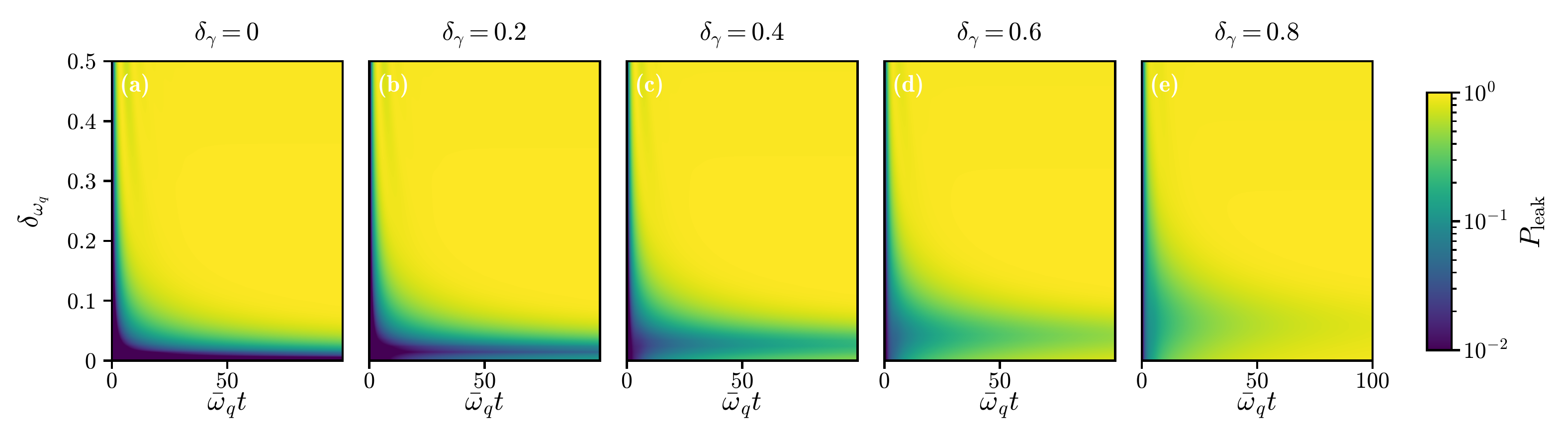}
    \caption{Singlet leakage $P_{\rm leak}(t)$ versus the relative qubit detuning $\delta_{\omega_q}$ for $\delta_\gamma=0$, $0.2$, $0.4$, $0.6$, and $0.8$. The mean frequency is fixed at $\bar{\omega}_q=\omega_\times$, and the logarithmic color scale spans $10^{-2}\leq P_{\rm leak}\leq1$.}
    \label{fig:qubit_detuning_heatmap}
\end{figure*}

We extend the analysis to the singlet-centered Werner family
\cite{werner_quantum_1989},
\begin{equation}
\hat{\rho}_{\mathrm W}(f)
=
f\,|\Psi^-\rangle\langle\Psi^-|
+
\frac{1-f}{3}
\left(
\hat{\mathbb I}_4-|\Psi^-\rangle\langle\Psi^-|
\right),
\label{eq:werner_state}
\end{equation}
where $f$ is the singlet fidelity. The maximally mixed state and pure singlet correspond to $f=1/4$ and $f=1$, respectively, while
$C(0,f)=\max\{0,2f-1\}$
\cite{wootters_entanglement_1998,hill_entanglement_1997}.

As shown in Fig.~\ref{fig:werner_concurrence}, the untwisted bath preserves the pure singlet and sustains concurrence in singlet-rich states, while also generating transient entanglement near $f=1/2$. Increasing $\delta_\gamma$ progressively suppresses these effects because the singlet, although dark at $\omega_\times$, couples to off-resonant spectral components. For $\delta_\gamma=0.8$, concurrence vanishes within the observation window for most of the Werner family.

\subsection{Combined effects of qubit detuning and spectral twisting}
\label{subsec:detun}

We introduce a relative detuning while fixing the mean qubit frequency at the spectral crossing:
\begin{equation}
\omega_{1,2}
=
\bar{\omega}_q(1\mp\delta_{\omega_q}),
\qquad
\bar{\omega}_q=\omega_\times ,
\label{eq:qubit_detuning_definition}
\end{equation}
so that $|\omega_2-\omega_1|=2\bar{\omega}_q\delta_{\omega_q}$. The system is initialized in $\ket{\Psi^-}$. We use $0\leq\delta_{\omega_q}\leq0.5$, $\sqrt{\gamma_1\gamma_2}=\omega_\times$, $\lambda_1=\lambda_2$, $J_{11}(\omega_\times)=J_{22}(\omega_\times)=0.1\,\bar{\omega}_q$, and $k_BT/(\hbar\bar{\omega}_q)=0.2$.

Restricted to the single-excitation subspace, the system Hamiltonian is
\begin{equation}
\left.\hat H_S\right|_{\mathcal H_1}
=
-\hbar\bar{\omega}_q\delta_{\omega_q}
\left(
\ket{\Psi^+}\bra{\Psi^-}
+
\ket{\Psi^-}\bra{\Psi^+}
\right),
\label{eq:detuning_BS_mixing}
\end{equation}
where $\ket{\Psi^+}=(\ket{10}+\ket{01})/\sqrt{2}$. Detuning therefore mixes the dark singlet with the bright symmetric state, removing exact protection even for the untwisted symmetric bath.

Figure~\ref{fig:qubit_detuning_heatmap} shows that spectral twisting further enhances the leakage. The two mechanisms are distinct: detuning mixes dark and bright sectors through $\hat H_S$, whereas twisting eliminates a frequency-independent dark coupling direction. Their combined action drives the leakage toward unity on progressively shorter timescales.

\section{Conclusions}
\label{sec:conclusions}

We have investigated the collective dynamics of two qubits coupled to a common bosonic environment through two correlated bath-force channels. In the maximally cross-spectrally coherent realization considered here, the matrix-valued spectral density $\mathbf J(\omega)$ has rank one wherever it is nonzero and therefore gives rise to one bright and one dark coupling direction at each frequency. 

A  \textit{global} dark channel exists only when the dark direction remains aligned over the relevant spectral interval $I$, that is when $ \displaystyle \cap_{\omega\in I}\ker\mathbf J(\omega) \neq \{\mathbf 0\}$.
Otherwise, the collective coupling direction rotates with frequency, and no single frequency-independent transformation diagonalizes the system–bath coupling over the entire spectrum. We term this geometric rotation \textit{spectral twisting} and quantify it through the Fubini--Study speed $\tau(\omega)$ of the bright spectral projector.

For transverse coupling, a pointwise dark coupling direction does not by itself imply a system state that is dark with respect to the full interaction. The bright coupling operator contains both lowering and raising components, $\hat Q_{\rm b}(\omega) = i \, [\hat L_{\rm b}(\omega)-\hat L_{\rm b}^\dagger(\omega)]$. At each frequency, \(\hat L_{\mathrm b}(\omega)\) annihilates a weighted single-excitation state \(\ket{D(\omega)}\), which is therefore locally lowering-dark. The same state is annihilated by $\hat L_{\rm b}^{\dagger}(\omega)$ only when the local spectral weights are equal, $J_{11}(\omega)=J_{22}(\omega)$. In that symmetric case, $\ket{D(\omega)}=\ket{\Psi^-}$ and is dark with respect to the full local coupling operator $\hat Q_{\rm b}(\omega)$. Exact dynamical protection further requires invariance of the dark subspace under the system Hamiltonian.

The central quantitative result is the weak-twisting expansion obtained for resonant qubits tuned to the spectral crossing $\omega_q=\omega_\times$. For spectral families with a fixed crossing and a bright spectral weight unchanged to linear order, the second-order leakage at fixed observation time satisfies
\begin{equation*}
P_{\rm leak}^{(2)}(t)
=
A(t)\bigl[\omega_\times\tau(\omega_\times)\bigr]^2
+
\mathcal O\!\left(
\bigl[\omega_\times\tau(\omega_\times)\bigr]^4
\right).
\end{equation*}

We tested this prediction for mismatched Drude--Lorentz spectra using hierarchical equations of motion generalized to cross-correlated bath forces. For this family, the relaxation-rate mismatch directly controls the weak-twisting parameter.

When both qubits are tuned to the spectral crossing $\omega_\times$, defined by \(J_{11}(\omega_\times)=J_{22}(\omega_\times)\), the singlet is locally dark at the transition frequency, but couples to off-resonant spectral components whose bright directions are rotated.  In the weak-twisting regime, at fixed observation time, we find quadratic scaling  of the leakage $P_{\mathrm{leak}}(t)\propto[\omega_\times\tau(\omega_\times)]^2$. Twisting also shifts the optimally protected state and accelerates the decay of Werner-state concurrence. Thus, identical values of $\mathbf J(\omega_q)$ can produce different dynamics when the off-resonant collective structures differ. In the secular GKLS description, the Kossakowski matrix is unchanged in this sweep. It remains rank one, so the singlet has no direct dissipative decay. Off-resonant twisting nevertheless enters the Lamb-shift Hamiltonian, which couples the singlet to the bright state, whose population then decays through the remaining dissipative channel.

The untwisted asymmetric control reveals a distinct limitation to dark-state protection. A frequency-independent collective direction guarantees a global lowering-dark state, but unequal local spectral weights generally prevent that
state from being annihilated by the corresponding raising operator.  Lowering darkness is sufficient in the zero-temperature rotating-wave limit, whereas finite-temperature absorption or counter-rotating processes can destroy this protection. Independently, qubit detuning produces leakage even for a symmetric environment by mixing dark and bright sectors through the system Hamiltonian.
Taken together, these results establish a hierarchy of
conditions for dark-state protection. The common-kernel and spectral-projector formulation developed here therefore provides a geometric framework for determining when correlated bosonic environments admit globally protected collective states. Extensions to frequency-dependent correlation phases, higher-rank spectral density matrices, and multiqubit systems are natural directions for future work.

\begin{acknowledgements}
 This work was supported by the Italian Ministry of University and Research through the PNRR MUR Project No. PE0000023-NQSTI.
\end{acknowledgements}

\newpage
\appendix
\section{Derivation of HEOM for cross-correlated bath channels}

\label{app:heom_derivation}

The hierarchical equations of motion (HEOM), originally introduced by Tanimura and Kubo \cite{tanimura_time_1989}, provide a nonperturbative framework for describing the reduced dynamics of open quantum systems coupled to environments with finite memory. The method maps the non-Markovian evolution of the reduced density operator onto a hierarchy of coupled equations involving a set of auxiliary density operators (ADOs),
which encode progressively higher-order system--bath memory effects. For a Gaussian environment linearly coupled to the system, the construction is based on an exponential representation of the bath correlation function. For certain spectral densities, including the Drude--Lorentz and
underdamped Brownian-oscillator forms, such decompositions can be obtained analytically, while more general correlation functions may be represented to controlled accuracy using, for example, Matsubara, Pad\'e, or numerical rational approximations \cite{lambert_qutip-bofin_2023}. The reduced dynamics is recovered upon convergence with respect to both the correlation-function decomposition and the hierarchy depth. In this sense, HEOM provides a numerically exact treatment without invoking the Born, Markov, secular, or rotating-wave approximations.

\subsection{System--bath Hamiltonian and assumptions}

The Hamiltonian of an open quantum system coupled to a bosonic reservoir is written as
\begin{equation}
    \hat H(t) = \hat H_S(t)+\hat H_B+\hat H_I ,
    \label{eq:heom_total_hamiltonian}
\end{equation}
where \(\hat H_S(t)\) is the bare Hamiltonian of the system and
\begin{equation}
    \hat H_B = \sum_k \hbar\omega_k \hat c_k^\dagger \hat c_k
\end{equation}
is the bare Hamiltonian of the bosonic reservoir, with $ [\hat c_k,\hat c_{k'}^\dagger] = \delta_{kk'}$. The interaction is assumed to be bilinear,
\begin{equation}
    \hat H_I = \hbar \sum_{\alpha}\hat Q_\alpha\otimes\hat B_\alpha ,
    \label{eq:heom_bilinear_interaction}
\end{equation}
where the operators \(\hat Q_\alpha\) act on the system Hilbert space and the operators \(\hat B_\alpha\) act on the reservoir Hilbert space. Both are assumed to be Hermitian; non-Hermitian couplings must be introduced together with their Hermitian-conjugate partners. The bath operators are linear combinations of the bosonic creation and annihilation operators. The index \(\alpha\) labels different system coupling channels interacting with the same reservoir and therefore allows for correlations between distinct channels.

We work in the interaction picture generated by
\begin{equation}
    \hat H_0(t)
    =
    \hat H_S(t)+\hat H_B,
\end{equation}
denoting by \(\overline O\) the interaction-picture counterpart of an operator \(\hat O\). Since the system and reservoir Hamiltonians act on different Hilbert spaces, the corresponding propagator factorizes as $ \hat U_0(t,0) = \hat U_S(t,0)\otimes \hat U_B(t), $ with
\begin{align}
    \hat U_S(t,0)
    &=
    \mathcal T
    \exp\left[
        -\frac{i}{\hbar}
        \int_0^t d\tau\,\hat H_S(\tau)
    \right],
    \\
    \hat U_B(t)
    &=
    \exp\left(
        -\frac{i}{\hbar}\hat H_B t
    \right),
\end{align}
where $\mathcal{T}$ denotes the time ordering operator. The total density operator in the interaction picture is given by
\begin{equation}
    \bar{\rho}_{\mathrm{tot}}(t)
    =
    \hat U_0^\dagger(t,0)
    \hat\rho_{\mathrm{tot}}(t)
    \hat U_0(t,0),
\end{equation}
and the interaction Hamiltonian becomes
\begin{equation}
    \bar H_I(t)
    =
    \hbar \sum_\alpha
    \bar Q_\alpha(t)\otimes\bar B_\alpha(t),
    \label{eq:interaction_picture_coupling}
\end{equation}
with $ \bar Q_\alpha(t) = \hat U_S^\dagger(t,0) \hat Q_\alpha \hat U_S(t,0)$ and $ \bar B_\alpha(t) = e^{i\hat H_Bt/\hbar} \hat B_\alpha e^{-i\hat H_Bt/\hbar}$. The interaction-picture Liouville--von Neumann equation reads
\begin{equation}
    \frac{d}{dt}\bar\rho_{\mathrm{tot}}(t)
    =
    -\bar{\mathcal L}_I(t)
    \bar\rho_{\mathrm{tot}}(t),
    \label{eq:interaction_picture_lvne}
\end{equation}
where
\begin{equation}
    \bar{\mathcal L}_I(t)\hat X
    \equiv
     \frac{i}{\hbar}
    \bar H_I^\times(t)\hat X.
    \label{eq:Liouvillian}
\end{equation}
and
\begin{equation}
    \hat A^\times\hat X
    \equiv
    [\hat A,\hat X].
\end{equation}
We assume: i) an initially factorized quantum state,
\begin{equation}
    \hat\rho_{\mathrm{tot}}(0)
    =
    \hat\rho_S(0)\otimes\hat\rho_B;
    \label{eq:factorized_initial_condition}
\end{equation}
ii) a stationary reservoir \([\hat\rho_B,\hat H_B]=0\); iii) a reservoir initially in the thermal quantum state \(\hat\rho_B=e^{-\beta\hat H_B}/\Tr_B\{e^{-\beta\hat H_B}\}\); iv) vanishing first moments for the reservoir operators,
\begin{equation}
\langle\hat B_\alpha\rangle_B=\Tr_B\{\hat B_\alpha\hat\rho_B\}=0 .
\label{eq:zero_bath_mean}
\end{equation}
The formal solution of Eq. \eqref{eq:interaction_picture_lvne}, traced over the bath Hilbert space, gives the reduced density operator
\begin{equation}
    \bar\rho_S(t)
    =
    \Tr_B
    \left\{
        \mathcal T
        \exp\left[
            -\int_0^t d\tau\,
            \bar{\mathcal L}_I(\tau)
        \right]
        \hat\rho_S(0)\otimes\hat\rho_B
    \right\}.
    \label{eq:formal_reduced_propagator}
\end{equation}

\subsection{Ordered cumulant expansion}

At this stage, some care is required. Consider first a classical zero-mean Gaussian stochastic process \(X(t)\), where \(X(t)\) is a commuting \(c\)-number random variable at each time and \(\langle\cdots\rangle\) denotes the ensemble average over its realizations. The condition of Gaussianity implies that all statistical moments are determined by the two-time covariance \(\langle X(t_1)X(t_2)\rangle\), while all cumulants of order higher than two vanish. Consequently, we have
\begin{multline}
    \left\langle
        \exp\left[
            \int_0^t d\tau\,X(\tau)
        \right]
    \right\rangle
    = \\
    \exp\left[
        \frac{1}{2}
        \int_0^t d\tau_1
        \int_0^t d\tau_2\,
        \langle X(\tau_1)X(\tau_2)\rangle
    \right].
    \label{eq:classical_gaussian_identity}
\end{multline}

Relation \eqref{eq:classical_gaussian_identity} cannot be applied directly to the interaction Liouvillian \(-\overline{\mathcal L}_I(t)\): the bath operators, and the associated system superoperators, do not commute at different times, so the exponential in Eq. \eqref{eq:formal_reduced_propagator} cannot be manipulated as an ordinary exponential of a commuting variable. The time ordering must therefore be retained throughout. Here and in what follows, \(\mathcal T\) is understood as ordering superoperators in Liouville space according to their time argument, placing later times to the left.

The appropriate generalization is Kubo's ordered (or partial) cumulant expansion \cite{kubo_generalized_1962,tanimura_numerically_2020}, in which the ordering symbol is kept in front of the exponential and the average is reorganized into ordered cumulants,
\begin{widetext}
\begin{equation}
\Tr_B\left\{
\mathcal T
\exp\left[
-\int_0^t d\tau\,\overline{\mathcal L}_I(\tau)
\right]
\hat\rho_S(0)\otimes\hat\rho_B
\right\}
= \\
\mathcal T
\exp\left[
\sum_{n=1}^{\infty}
\frac{(-1)^{n}}{n!}
\int_0^t dt_1\cdots\int_0^t dt_n\,
\left\langle\!\left\langle
\overline{\mathcal L}_I(t_1)\cdots\overline{\mathcal L}_I(t_n)
\right\rangle\!\right\rangle_{B,\mathrm{oc}}
\right]
\hat\rho_S(0),
\label{eq:ordered_cumulant_expansion}
\end{equation}
\end{widetext}
where $\langle\!\langle\cdots\rangle\!\rangle_{B,\mathrm{oc}}$ denotes the ordered bath cumulant, defined by the usual cumulant recursion applied to the time-ordered products. The Gaussian character of the bath implies that all connected time-ordered bath correlation functions beyond second order vanish. Equivalently, Wick's theorem reduces every higher-order bath moment to a sum over products of two-point correlation functions. Because the system superoperators are not averaged over, their noncommutativity is retained through the global time-ordering operator. Resummation of the resulting pair contractions therefore yields an exact influence superoperator that is quadratic in the system coupling superoperators.

Keeping the surviving term, the reduced propagator is
\begin{equation}
\overline{\mathcal U}_S(t)
=
\mathcal T
\exp\left\{
\frac{1}{2}
\int_0^t dt_1
\int_0^t dt_2\,
\left\langle\!\left\langle
\overline{\mathcal L}_I(t_1)
\overline{\mathcal L}_I(t_2)
\right\rangle\!\right\rangle_{B,\mathrm{oc}}
\right\}.
\label{eq:ordered_second_cumulant}
\end{equation}
Using the definition Eq. \eqref{eq:Liouvillian} the second ordered cumulant acts on a system operator as
\begin{multline}
\left\langle\!\left\langle
\overline{\mathcal L}_I(t_1)
\overline{\mathcal L}_I(t_2)
\right\rangle\!\right\rangle_{B,\mathrm{oc}}
\bullet
\\
=
-\frac{1}{\hbar^2}
\Tr_B\left\{
\mathcal T\,
\overline H_I^{\times}(t_1)
\overline H_I^{\times}(t_2)
\left[
\bullet\otimes\hat\rho_B
\right]
\right\}.
\label{eq:ordered_liouvillian_cumulant}
\end{multline}
Substitution into Eq. \eqref{eq:ordered_second_cumulant} then gives
\begin{multline}
\overline{\mathcal U}_S(t)
=
\mathcal T
\exp\Biggl\{
-\frac{1}{2\hbar^2}
\int_0^t dt_1
\int_0^t dt_2
\\
\times
\Tr_B\left\{
\mathcal T\,
\overline H_I^{\times}(t_1)
\overline H_I^{\times}(t_2)
\left[
\bullet\otimes\hat\rho_B
\right]
\right\}
\Biggr\}.
\label{eq:full_square_influence}
\end{multline}
Precisely because of the ordering operator, the integrand of Eq. \eqref{eq:full_square_influence} is symmetric under \(t_1\leftrightarrow t_2\). The integral over the square \(0\leq t_1,t_2\leq t\) may therefore be replaced by twice the integral over the ordered triangle \(0\leq t_2\leq t_1\leq t\), on which the ordering is explicit and \(\mathcal T\) can be dropped from the integrand,
\begin{multline}
\overline{\mathcal U}_S(t)
=
\mathcal T
\exp\Biggl\{
-\frac{1}{\hbar^2}
\int_0^t dt_1
\int_0^{t_1} dt_2
\\
\times
\Tr_B\left\{
\overline H_I^{\times}(t_1)
\overline H_I^{\times}(t_2)
\left[
\bullet\otimes\hat\rho_B
\right]
\right\}
\Biggr\}.
\label{eq:triangular_influence}
\end{multline}

\subsection{Bath correlation matrix and influence phase}

We define the matrix of bath correlation functions
\begin{equation}
    C_{\alpha\beta}(t-s)
    =
    \Tr_B
    \left\{
        \bar B_\alpha(t)
        \bar B_\beta(s)
        \hat\rho_B
    \right\}.
    \label{eq:bath_correlation_matrix}
\end{equation}
which by stationarity depends only on the time difference. For Hermitian bath coupling operators,
\begin{equation}
C_{\beta\alpha}(-\tau)=C_{\alpha\beta}^{*}(\tau).
\label{eq:correlation_hermiticity}
\end{equation}
Writing \(C_{\alpha\beta}(t)=C^{\mathrm R}_{\alpha\beta}(t) +iC^{\mathrm I}_{\alpha\beta}(t)\) with real \(C^{\mathrm R}_{\alpha\beta}\) and \(C^{\mathrm I}_{\alpha\beta}\), Eq. \eqref{eq:correlation_hermiticity} becomes
\begin{equation}
C^{\mathrm R}_{\beta\alpha}(-\tau)=C^{\mathrm R}_{\alpha\beta}(\tau),
\qquad
C^{\mathrm I}_{\beta\alpha}(-\tau)=-C^{\mathrm I}_{\alpha\beta}(\tau).
\label{eq:correlation_parity}
\end{equation}

Performing the Gaussian bath trace gives the exact reduced propagator
\begin{equation}
\overline\rho_S(t)
= \mathcal T e^{-\overline{\Phi}(t)}
\hat\rho_S(0).
\label{eq:influencefunctionalform}
\end{equation}
where \(\overline\Phi(t)\) is the influence phase
\begin{equation}
\overline{\Phi}(t)
=
\int_0^t dt_1
\int_0^{t_1}dt_2\,
\overline{\mathcal K}(t_1,t_2).
\end{equation}
and the kernel reads
\begin{multline}
\overline{\mathcal K}(t_1,t_2)
=
\sum_{\alpha,\beta=1}^{2}
\overline Q_\alpha^\times(t_1)
\Big[
C_{\alpha\beta}^{\mathrm R}(t_1-t_2)
\overline Q_\beta^\times(t_2)
\\
+
iC_{\alpha\beta}^{\mathrm I}(t_1-t_2)
\overline Q_\beta^\circ(t_2)
\Big].
\label{eq:Kernel}
\end{multline}
with the anticommutator superoperator \(\hat A^\circ\hat X\equiv\{\hat A,\hat X\}\). The real part of the correlation function describes fluctuations, whereas the imaginary part encodes the dissipative response and the bath-induced renormalization of the system dynamics. Equation \eqref{eq:influencefunctionalform} is exact under the assumptions of an initially factorized state, a Gaussian bath and a bilinear system--bath interaction; no Born, Markov, secular or rotating-wave approximation has been introduced.

\subsection{Exponential representation and hierarchy construction}

For $t\geq0$, we represent the real and imaginary parts of each correlation function by finite exponential expansions,
\begin{subequations}
\begin{align}
C_{\alpha\beta}^{\mathrm R}(t)
&\simeq
\sum_{k=1}^{N_{\alpha\beta}^{R}}
c_{\alpha\beta,k}^{R}
e^{-\gamma_{\alpha\beta,k}^{R}t},
\\
C_{\alpha\beta}^{\mathrm I}(t)
&\simeq
\sum_{k=1}^{N_{\alpha\beta}^{I}}
c_{\alpha\beta,k}^{I}
e^{-\gamma_{\alpha\beta,k}^{I}t}.
\end{align}
\label{eq:Corr_func_decomposition}
\end{subequations}
The representation may be exact for particular correlation functions or may be obtained numerically to arbitrary accuracy. The exact reduced dynamics is recovered upon convergence with respect to both the correlation-function decomposition and the hierarchy depth.

For two correlated coupling channels, the correlation function is matrix valued:

\begin{equation}
\mathbf C(t)
=
\begin{pmatrix}
C_{11}(t) & C_{12}(t) \\
C_{21}(t) & C_{22}(t)
\end{pmatrix}
\end{equation}
Inserting Eq. \eqref{eq:Corr_func_decomposition} into Eq. \eqref{eq:Kernel}, the kernel becomes
\begin{widetext}
\begin{equation}
 \overline{\mathcal{K}}(t_1, t_2) =  \sum_{\alpha,\beta=1}^{2}
\bar{Q}_{\alpha}^{\times}(t_1)
\Bigg[
\sum_{k=1}^{\Nr}
c_{\alpha\beta,k}^R
e^{-\gamma_{\alpha\beta,k}^R (t_1-t_2)}
\bar{Q}_{\beta}^{\times}(t_2)
 \\ + i
\sum_{k=1}^{N_{\alpha\beta}^I}
c_{\alpha\beta,k}^I
e^{-\gamma_{\alpha\beta,k}^I (t_1-t_2)}
\bar{Q}_{\beta}^{\circ}(t_2)
\Bigg]
\label{eq:KernelExpansion}
\end{equation}
\end{widetext}
Each exponential component of each matrix element thus defines a distinct memory channel, labelled by the triple \((\alpha,\beta,k)\) together with the real or imaginary character of the corresponding correlation component. The hierarchy is built by promoting each of these channels to an independent dynamical variable.

\subsubsection{First-tier auxiliary operators}
Returning to the Schr\"odinger picture, Eq. \eqref{eq:influencefunctionalform} reads
\begin{equation}
    \hat{\rho}_S(t)
    = U_S(t,0) \timeord e^{-\overline{\Phi}(t)} \hat{\rho}_S(0) U_S^\dagger(t,0).
    \label{eq:SchrodingerPix}
\end{equation}
Differentiating Eq. \eqref{eq:SchrodingerPix} and using the Dyson identity for time-ordered exponentials, we obtain
\begin{multline}
\dot{\hat\rho}_S(t)
=
-\frac{i}{\hbar}
\hat H_S^\times(t)\hat\rho_S(t)
\\
+
\hat U_S(t,0)
\mathcal T
\left\{
-\int_0^t d\tau\,
\overline{\mathcal K}(t,\tau)
e^{-\overline{\Phi}(t)}
\right\}
\hat\rho_S(0)
\hat U_S^\dagger(t,0).
\label{eq:dotRho}
\end{multline}
where the second term collects the memory contributions accumulated over the whole history \(0\leq\tau\leq t\). Substituting Eq. \eqref{eq:KernelExpansion} into Eq. \eqref{eq:dotRho} we obtain
\begin{widetext}
\begin{multline}
\dot{\hat{\rho}}_S(t)
=
-\frac{i}{\hbar}
\hat H_S(t)^\times\hat{\rho}_S(t)
-
\hat U_S(t,0)
\sum_{\alpha,\beta=1}^{2}
\sum_{k=1}^{N_{\alpha\beta}^{R}}
c_{\alpha\beta,k}^{R}
\Bigg[
\bar Q_\alpha^\times(t)
\mathcal T
\left\{
\int_0^t d\tau\,
e^{-\gamma_{\alpha\beta,k}^{R}(t-\tau)}
\bar Q_\beta^\times(\tau)
e^{-\overline{\Phi}(t)}
\right\}
\hat\rho_S(0)
\Bigg]
\hat U_S^\dagger(t,0)
\\
-i \, \hat U_S(t,0)
\sum_{\alpha,\beta=1}^{2}
\sum_{k=1}^{N_{\alpha\beta}^{I}}
c_{\alpha\beta,k}^{I}
\Bigg[
\bar Q_\alpha(t)^\times
\mathcal T
\left\{
\int_0^t d\tau\,
e^{-\gamma_{\alpha\beta,k}^{I}(t-\tau)}
\bar Q_\beta^\circ(\tau)
e^{-\overline{\Phi}(t)}
\right\}
\hat\rho_S(0)
\Bigg]
\hat U_S^\dagger(t,0).
\label{eq:derivative_reduced_density}
\end{multline}
\end{widetext}

Two manipulations bring Eq. \eqref{eq:derivative_reduced_density} to a closed form. First, since \(\overline Q_\alpha^\times(t)\) is evaluated at the latest time, the time-ordering operator places it to the left of all superoperators evaluated at earlier times, and it may be extracted from \(\mathcal T\). Second, it is transformed to the Schr\"odinger picture by unitary transformation with the free evolution superoperator, using the covariance of commutators under conjugation,
\begin{equation}
\hat U_S(t,0)
\left[
\overline Q_\alpha^\times(t)\,\overline X
\right]
\hat U_S^\dagger(t,0)
=
\hat Q_\alpha^\times
\left[
\hat U_S(t,0)\,
\overline X\,
\hat U_S^\dagger(t,0)
\right]
\label{eq:commutator_covariance}
\end{equation}
for any interaction-picture system operator $\overline X$. Eventually, we obtain
\begin{widetext}
\begin{multline}
    \dot{\hat{\rho}}_S(t)
    = -\frac{i}{\hbar} \hat{H}_S^\times(t) \hat{\rho}_S(t) - \sum_{\alpha,\beta=1}^{2} \sum_{k=1}^{\Nr} \ckR \hat{Q}_\alpha^{\times}        \hat{U}_S(t,0) \timeord \left\{
      \int_0^t d \tau \,   e^{-\gammakR (t - \tau)}
    \,\bar{Q}_\beta(\tau)^{\times}  e^{-\overline{\Phi}(t)}  \right\} \hat{\rho}_S(0) \hat{U}_S^\dagger(t,0)   \\ - i \sum_{\alpha,\beta=1}^{2}  \sum_{k=1}^{\Ni} \ckI \hat{Q}_\alpha^{\times} \hat{U}_S(t,0)    \timeord \left\{
       \int_0^t d \tau \, e^{-\gammakI (t - \tau)}
    \,\bar{Q}_\beta(\tau)^{\circ}  e^{-\overline{\Phi}(t)}  \right\} \hat{\rho}_S(0) \hat{U}_S^\dagger(t,0)
    \label{eq:DerivativeRhoS}
\end{multline}
\end{widetext}
Every memory contribution in Eq. \eqref{eq:DerivativeRhoS} now has the same structure: a Schr\"odinger-picture commutator \(\hat Q_\alpha^\times\) acting from the left on an object that is itself an exponentially weighted history integral. It is therefore natural to introduce the interaction-picture memory superoperators.
\begin{subequations}
\begin{align}
\overline{\mathcal B}_{\alpha\beta,k}^{R}(t)
&=
-i
\int_{0}^{t}d\tau\,
c_{\alpha\beta,k}^{R}
e^{-\gamma_{\alpha\beta,k}^{R}(t-\tau)}
\overline Q_{\beta}^{\times}(\tau),
\\
\overline{\mathcal B}_{\alpha\beta,k}^{I}(t)
&=
+
\int_{0}^{t}d\tau\,
c_{\alpha\beta,k}^{I}
e^{-\gamma_{\alpha\beta,k}^{I}(t-\tau)}
\overline Q_{\beta}^{\circ}(\tau).
\end{align}
\label{eq:memory_superoperators}
\end{subequations}
one for each memory channel, and to define the corresponding first-tier auxiliary density operators (ADOs) $\hat{\rho}^{\mathbf e_{\alpha\beta,k}^{R}}(t)$, with $k=1,\ldots,N_{\alpha\beta}^{R}$, and $\hat{\rho}^{\mathbf e_{\alpha\beta,k}^{I}}(t)$, with $k=1,\ldots,N_{\alpha\beta}^{I}$, as
\begin{subequations}
\begin{align}
\hat{\rho}^{\mathbf e_{\alpha\beta,k}^{R}}(t)
&=
\hat U_S(t,0)\,
\mathcal T
\left\{
\overline{\mathcal B}_{\alpha\beta,k}^{R}(t)
e^{-\overline{\Phi}(t)}
\right\}
\hat\rho_S(0)\,
\hat U_S^\dagger(t,0),
\\
\hat{\rho}^{\mathbf e_{\alpha\beta,k}^{I}}(t)
&=
\hat U_S(t,0)\,
\mathcal T
\left\{
\overline{\mathcal B}_{\alpha\beta,k}^{I}(t)
e^{-\overline{\Phi}(t)}
\right\}
\hat\rho_S(0)\,
\hat U_S^\dagger(t,0).
\end{align}
\label{eq:first_tier_ADOs}
\end{subequations}
The unit multi-indices \(\mathbf e_{\alpha\beta,k}^{R}\) and \(\mathbf e_{\alpha\beta,k}^{I}\) are defined by \begin{subequations}
\begin{align}
\left(\mathbf e_{\alpha\beta,k}^{R}\right)_{\mu\nu,l}^{R} &= \delta_{\alpha\mu}\delta_{\beta\nu}\delta_{kl}, & \left(\mathbf e_{\alpha\beta,k}^{R}\right)_{\mu\nu,l}^{I} &= 0, \\
\left(\mathbf e_{\alpha\beta,k}^{I}\right)_{\mu\nu,l}^{I} &= \delta_{\alpha\mu}\delta_{\beta\nu}\delta_{kl}, & \left(\mathbf e_{\alpha\beta,k}^{I}\right)_{\mu\nu,l}^{R} &= 0.
\end{align}
\end{subequations} These first-tier ADOs are not physical density matrices: they are auxiliary quantities that store the portion of the system--bath correlation carried by one particular exponential component of one particular element of \(\mathbf C(t)\).

Inserting Eq. \eqref{eq:first_tier_ADOs} into Eq. \eqref{eq:DerivativeRhoS}, the equation of motion for the physical density operator takes the compact form
\begin{multline}
\dot{\hat{\rho}}_S(t)=-\frac{i}{\hbar} \hat{H}_S^\times(t) \hat{\rho}_S(t) \\ - i \sum_{\alpha,\beta=1}^{2} \hat Q_{\alpha}^{\times} \Bigg[ \sum_{k=1}^{N_{\alpha\beta}^{R}} \hat{\rho}^{\mathbf e_{\alpha\beta,k}^{R}}(t) + \sum_{k=1}^{N_{\alpha\beta}^{I}} \hat{\rho}^{\mathbf e_{\alpha\beta,k}^{I}}(t) \Bigg].
\label{eq:HEOM_physical_matrix}
\end{multline}
Equation \eqref{eq:HEOM_physical_matrix} is exact, but not closed: the time derivative of each first-tier ADO generates, through the derivative of \(e^{-\overline\Phi(t)}\), a further factor \(\overline{\mathcal B}_{\alpha\beta,k}^{R,I}\), that is, an object carrying two memory factors. Closing the system therefore requires operators carrying an arbitrary number of memory factors of each channel, which is precisely the hierarchy constructed below.

\subsubsection{Arbitrary-tier hierarchy}
\begin{figure}[t]
    \centering
    \includegraphics[width=\linewidth]{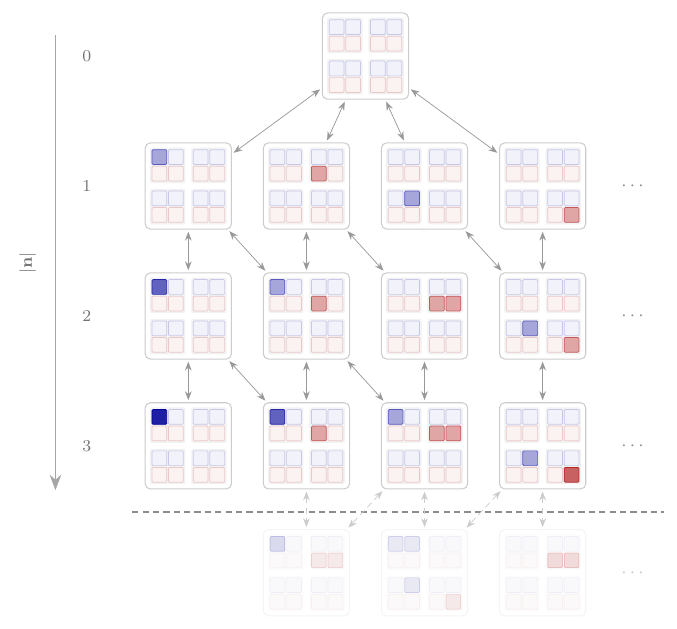}
\caption{Structure of the HEOM hierarchy for two cross-correlated coupling
channels, with $N^{R}_{\alpha\beta}=N^{I}_{\alpha\beta}=2$ exponential
components per element of $\mathbf{C}(t)$ and hence $M=16$ memory channels. ADOs are
organized by tier $\ell=|\mathbf{n}|$, starting from the root
$\hat\rho^{\mathbf{0}}=\hat\rho_{S}$; darker fill indicates higher
occupation. Double-headed arrows denote the couplings to
$\mathbf{n}\pm\mathbf{e}^{R,I}_{\alpha\beta,k}$ in Eq.~\eqref{eq:HEOM_complete}.
Only representative ADOs are shown at each tier. The dashed line marks the
truncation at $N_{C}=3$, beyond which the faded nodes and the couplings
reaching them are discarded.}
\label{fig:heom-hierarchy}
    \label{fig:multi_index}
\end{figure}

We therefore collect the occupation numbers of all memory channels into the multi-index
\begin{equation}
\mathbf n
=
\left(
\left\{
n_{\alpha\beta,k}^{R}
\right\}_{k=1}^{N_{\alpha\beta}^{R}},
\left\{
n_{\alpha\beta,k}^{I}
\right\}_{k=1}^{N_{\alpha\beta}^{I}}
\right)_{\alpha,\beta=1,2},
\label{eq:multi-index}
\end{equation}
where every component $n_{\alpha\beta,k}^{R,I}$ is a non-negative integer. The structure of \(\mathbf n\) is illustrated in Fig. \ref{fig:multi_index}.

The generic interaction-picture ADO is then defined as
\begin{multline}
\overline{\rho}^{\mathbf n}(t)
=
\mathcal T
\Bigg\{
\prod_{\alpha,\beta}
\prod_{k=1}^{N_{\alpha\beta}^{R}}
\left[
\overline{\mathcal B}_{\alpha\beta,k}^{R}(t)
\right]^{n_{\alpha\beta,k}^{R}}
\\
\times
\prod_{\alpha,\beta}
\prod_{k=1}^{N_{\alpha\beta}^{I}}
\left[
\overline{\mathcal B}_{\alpha\beta,k}^{I}(t)
\right]^{n_{\alpha\beta,k}^{I}}
e^{-\overline{\Phi}(t)}
\Bigg\}
\hat\rho_S(0),
\label{eq:generic_interaction_picture_ADO}
\end{multline}
and its Schr\"odinger-picture counterpart is
\begin{equation}
\hat\rho^{\mathbf n}(t)
=
\hat U_S(t,0)\,
\overline{\rho}^{\mathbf n}(t)\,
\hat U_S^\dagger(t,0).
\label{eq:generic_schrodinger_picture_ADO}
\end{equation}
In Eq. \eqref{eq:generic_interaction_picture_ADO}, the time-ordering operator acts globally on all elementary system superoperators contained in the memory superoperators \(\overline{\mathcal B}_{\alpha\beta,k}^{R,I}(t)\) and in the influence functional.

The physical reduced density operator corresponds to the zeroth-tier element, \( \hat\rho_S(t)=\hat\rho^{\mathbf 0}(t) \). Differentiating Eq. \eqref{eq:generic_schrodinger_picture_ADO} and repeating the two manipulations used above produces the multiplicity factors \(n_{\alpha\beta,k}^{R,I}\), the damping terms proportional to \(\gamma_{\alpha\beta,k}^{R,I}\), and the couplings to the neighbouring tiers \(\mathbf n\pm\mathbf e_{\alpha\beta,k}^{R,I}\):
\begin{widetext}
\begin{multline}
\dot{\hat{\rho}}^{\mathbf n}(t)
=
\mathcal L_S(t) \hat{\rho}^{\mathbf n}(t)
-
\sum_{\alpha,\beta=1}^{2}
\left(
\sum_{k=1}^{N_{\alpha\beta}^{R}}
n_{\alpha\beta,k}^{R}\gamma_{\alpha\beta,k}^{R}
+
\sum_{k=1}^{N_{\alpha\beta}^{I}}
n_{\alpha\beta,k}^{I}\gamma_{\alpha\beta,k}^{I}
\right)
\hat{\rho}^{\mathbf n}(t)
-i
\sum_{\alpha,\beta=1}^{2}
\sum_{k=1}^{N_{\alpha\beta}^{R}}
n_{\alpha\beta,k}^{R}
c_{\alpha\beta,k}^{R}
\hat Q_{\beta}^{\times}
\hat{\rho}^{\mathbf n-\mathbf e_{\alpha\beta,k}^{R}}(t)
\\
+
\sum_{\alpha,\beta=1}^{2}
\sum_{k=1}^{N_{\alpha\beta}^{I}}
n_{\alpha\beta,k}^{I}
c_{\alpha\beta,k}^{I}
\hat Q_{\beta}^{\circ}
\hat{\rho}^{\mathbf n-\mathbf e_{\alpha\beta,k}^{I}}(t)
-i
\sum_{\alpha,\beta=1}^{2}
\hat Q_{\alpha}^{\times}
\left[
\sum_{k=1}^{N_{\alpha\beta}^{R}}
\hat{\rho}^{\mathbf n+\mathbf e_{\alpha\beta,k}^{R}}(t)
+
\sum_{k=1}^{N_{\alpha\beta}^{I}}
\hat{\rho}^{\mathbf n+\mathbf e_{\alpha\beta,k}^{I}}(t)
\right].
\label{eq:HEOM_complete}
\end{multline}
\end{widetext}

The first term describes the free system dynamics generated by
\begin{equation}
\mathcal L_S(t)\,\bullet
\equiv
-\frac{i}{\hbar}
\hat H_S^\times(t)\,\bullet .
\label{eq:system_liouvillian}
\end{equation}
The second term results from differentiating the exponential memory kernels. Each occurrence of a real or imaginary memory channel contributes its corresponding decay rate, producing the multiplicity factors \(n_{\alpha\beta,k}^{R}\) and \(n_{\alpha\beta,k}^{I}\). The third and fourth terms arise from the upper integration limits of the real and imaginary memory integrals, respectively. They remove one occurrence of the corresponding memory channel and therefore couple \(\hat{\rho}^{\mathbf n}(t)\) to the lower-tier ADOs \(\hat{\rho}^{\mathbf n-\mathbf e_{\alpha\beta,k}^{R}}(t)\) and \(\hat{\rho}^{\mathbf n-\mathbf e_{\alpha\beta,k}^{I}}(t)\). Finally, differentiation of the influence functional generates the last term, which introduces an additional memory factor and couples the ADO to the neighboring upper tier \(\mathbf n+\mathbf e_{\alpha\beta,k}^{R,I}\).

For the factorized initial condition in Eq. \eqref{eq:factorized_initial_condition}, the hierarchy is initialized according to
\begin{equation}
\hat\rho^{\mathbf 0}(0)=\hat\rho_S(0),
\qquad
\hat\rho^{\mathbf n}(0)=0
\quad
\text{for }
\mathbf n\neq\mathbf 0.
\label{eq:HEOM_initial_conditions}
\end{equation}

\subsection{Auxiliary-operator count for a uniform tier cutoff}

\label{sec:ADOsHEOM}
In the present correlated-bath construction, each auxiliary density operator is labeled by the occupation numbers associated with the real and imaginary exponential components of every correlation-matrix element, according to Eq. \eqref{eq:multi-index}. The total number of independent hierarchy indices is therefore
\begin{equation}
M
=
\sum_{\alpha,\beta=1}^{2}
\left(
N_{\alpha\beta}^{R}
+
N_{\alpha\beta}^{I}
\right).
\label{eq:number_heom_modes}
\end{equation}
The tier of an ADO is defined as the total occupation number
\begin{equation}
\ell
=
|\mathbf n|
=
\sum_{\alpha,\beta=1}^{2}
\left[
\sum_{k=1}^{N_{\alpha\beta}^{R}}
n_{\alpha\beta,k}^{R}
+
\sum_{k=1}^{N_{\alpha\beta}^{I}}
n_{\alpha\beta,k}^{I}
\right].
\label{eq:heom_tier_definition}
\end{equation}
For a fixed tier \(\ell\), the number of non-negative integer multi-indices satisfying \(|\mathbf n|=\ell\) is
\begin{equation}
N_{\mathrm{ADO}}(\ell)
=
\binom{M+\ell-1}{\ell}.
\label{eq:ado_number_fixed_tier}
\end{equation}
If the hierarchy is truncated uniformly at the maximum tier \(N_C\), namely by retaining all ADOs satisfying \(|\mathbf n|\leq N_C\), the total number of retained operators is
\begin{equation}
N_{\mathrm{ADO}}(\leq N_C)
=
\sum_{\ell=0}^{N_C}
\binom{M+\ell-1}{\ell}
=
\binom{M+N_C}{N_C}.
\label{eq:ado_number_uniform_cutoff}
\end{equation}
This count includes the zeroth-tier ADO \(\hat\rho^{\mathbf 0}(t)=\hat\rho_S(t)\), corresponding to the physical reduced density operator.

If the same numbers of exponential terms are used for every matrix element,
\begin{equation}
N_{\alpha\beta}^{R}=N_R,
\qquad
N_{\alpha\beta}^{I}=N_I,
\end{equation}
then, because \(\alpha,\beta\in\{1,2\}\), the number of hierarchy modes reduces to
\begin{equation}
M=4(N_R+N_I).
\label{eq:uniform_number_heom_modes}
\end{equation}

\section{Validation of the cross-correlated HEOM}\label{app:HEOM_validation}
In this section, we consider two examples to validate the HEOM model described in Appendix \ref{app:heom_derivation}. Specifically, we compare its results with those obtained from the Gorini--Kossakowski--Sudarshan--Lindblad master equation \cite{gorini_completely_1976,breuer_theory_2007} for two correlated coupling channels and from an analytically solvable pure-dephasing model known as the Unruh model \cite{Unruh1995MaintainingCoherence}.

\subsection{Secular GKLS benchmark for correlated channels}
\label{sec:GKLS}
We derive the GKLS master equation for two correlated baths following the global approach of Refs. \cite{gorini_completely_1976,breuer_theory_2007}. The system coupling operators are decomposed as
\begin{equation}
    \hat A_\alpha(\omega) = \sum_{\epsilon'-\epsilon=\hbar\omega}\hat\Pi(\epsilon)\hat Q_\alpha\hat\Pi(\epsilon'),
    \label{eq:global_jump_decomposition}
\end{equation}
with
\begin{equation}
[\hat H_S,\hat A_\alpha(\omega)]=-\hbar \omega \hat A_\alpha(\omega),
    \qquad
\hat Q_\alpha=\sum_{\omega}\hat A_\alpha(\omega).
    \label{eq:eigenoperator_properties}
\end{equation}
For the Hamiltonian in Eq. \eqref{eq:H_sys}, with $\omega_1 = \omega_2 = \omega_q$, the relevant Bohr frequencies are $\omega=\pm\omega_q$. Using the raising and lowering operators $\hat\sigma_y=i(\hat\sigma_--\hat\sigma_+)$, one obtains
\begin{subequations}
\label{eq:A_omega_two_qubits}
\begin{align}
    \hat A_1(\omega_q)&=
    i\hat\sigma_-^{(1)},
    &
    \hat A_2(\omega_q)
    &=
    i\hat\sigma_-^{(2)},
    \\
    \hat A_1(-\omega_q)
    &=
    -i\hat\sigma_+^{(1)},
    &
    \hat A_2(-\omega_q)
    &=
    -i\hat\sigma_+^{(2)} .
\end{align}
\end{subequations}

We define the one-sided Fourier transform
\begin{equation}
    \mathcal G_{\alpha\beta}(\omega)
    =
    \int_0^\infty d\tau\,
    e^{i\omega\tau}
    C_{\alpha\beta}(\tau).
    \label{eq:one_sided_transform_corr}
\end{equation}

In the full secular approximation only the resonant terms $\omega=\omega'$ are retained and this gives the secular GKLS master equation
\begin{widetext}
\begin{multline}    
\frac{d}{dt}\rho_S(t)
    =
    -\frac{i}{\hbar}
    \left[
        \hat H_S+\hat H_{\rm LS},
        \rho_S(t)
    \right]
    \quad+
    \sum_{\omega=\pm\omega_q}
    \sum_{\alpha,\beta=1}^{2}
    \Gamma_{\alpha\beta}(\omega)
    \left[
        \hat A_\beta(\omega)
        \rho_S(t)
        \hat A_\alpha^\dagger(\omega)
        -
        \frac{1}{2}
        \left\{
            \hat A_\alpha^\dagger(\omega)
            \hat A_\beta(\omega),
            \rho_S(t)
        \right\}
    \right].
\end{multline}
\end{widetext}

The rate and Lamb-shift matrices are
\begin{subequations}
\begin{align}
    \Gamma_{\alpha\beta}(\omega)&=\mathcal G_{\alpha\beta}(\omega)+\mathcal G_{\beta\alpha}^{*}(\omega),
    \label{eq:rate_matrix_corr}
    \\
    S_{\alpha\beta}(\omega)&=\frac{\mathcal G_{\alpha\beta}(\omega)-\mathcal G_{\beta\alpha}^{*}(\omega)}{2i}.
    \label{eq:lamb_shift_matrix_corr}
\end{align}
\end{subequations}
The Lamb-shift Hamiltonian is
\begin{equation}
    \hat H_{\rm LS}=\sum_{\omega=\pm\omega_q}\sum_{\alpha,\beta=1}^{2}\hbar S_{\alpha\beta}(\omega)\hat A_\alpha^\dagger(\omega)\hat A_\beta(\omega).
    \label{eq:lamb_shift_hamiltonian_corr}
\end{equation}

Using the exponential decomposition of the correlation functions in Eq. \eqref{eq:Corr_func_decomposition} and applying the Fourier transform one obtains
\begin{equation}
    \mathcal G_{\alpha\beta}(\omega) =\sum_{k=1}^{N_{\alpha\beta}^R}\frac{c_{\alpha\beta,k}^R}{\gamma_{\alpha\beta,k}^R-i\omega}+i\sum_{k=1}^{N_{\alpha\beta}^I}\frac{c_{\alpha\beta,k}^I}{\gamma_{\alpha\beta,k}^I-i\omega}.
    \label{eq:one_sided_transform_exp}
\end{equation}

For each Bohr frequency, the rate matrix, also called the \emph{Kossakowski matrix} \cite{gorini_completely_1976,cattaneo2020symmetry}, is
\begin{equation}
    \Gamma(\omega)
    =
    \begin{pmatrix}
        \Gamma_{11}(\omega) & \Gamma_{12}(\omega) \\
        \Gamma_{21}(\omega) & \Gamma_{22}(\omega)
    \end{pmatrix},
    \qquad
    \Gamma_{12}(\omega)
    =
    |\Gamma_{12}(\omega)|e^{i\phi_\omega},
    \label{eq:gamma_matrix}
\end{equation}
with $\Gamma_{12}(\omega) = \Gamma_{21}^*(\omega)$.

The eigenvalues of the Kossakowski matrix in Eq. \eqref{eq:gamma_matrix} are
\begin{multline}
r_{\pm}(\omega) = \frac{\Gamma_{11}(\omega)+\Gamma_{22}(\omega)}{2}\
\\
\pm \sqrt{\left[\frac{\Gamma_{11}(\omega)-\Gamma_{22}(\omega)}{2}\right]^2+|\Gamma_{12}(\omega)|^2}.
\label{eq:gamma_eigenvalues_2x2}
\end{multline}
The corresponding eigenvectors can be parametrized through the frequency-dependent mixing angle
\begin{equation}
\tan\left(2\theta_\omega\right)
=
\frac{2|\Gamma_{12}(\omega)|}
{\Gamma_{11}(\omega)-\Gamma_{22}(\omega)},
\label{eq:theta_definition}
\end{equation}
where the branch of $\theta_\omega$ is chosen such that $\mathbf u_+(\omega)$ corresponds to $r_+(\omega)$. A convenient choice of normalized eigenvectors is
\begin{subequations}
\label{eq:gamma_eigenvectors_2x2}
\begin{align}
\mathbf u_+(\omega)
&=
(\cos\theta_\omega\  ,  e^{-i\phi_\omega}\sin\theta_\omega)^\mathrm{T},
\\
\mathbf u_-(\omega)
&=
(-\sin\theta_\omega\ , e^{-i\phi_\omega}\cos\theta_\omega)^\mathrm{T}.
\end{align}
\end{subequations}

The off-diagonal matrix element in the collective basis is
\begin{multline}
    \Gamma'_{+-}(\omega)
    \equiv
    \mathbf u_+^\dagger(\omega)
    \Gamma(\omega)
    \mathbf u_-(\omega)
    \\
    =
    |\Gamma_{12}(\omega)|
    \cos\left(2\theta_\omega\right)
    -
    \frac{
        \Gamma_{11}(\omega)-\Gamma_{22}(\omega)
    }{2}
    \sin\left(2\theta_\omega\right)
    =0,
    \label{eq:gamma_rotated_offdiagonal}
\end{multline}

where Eq. \eqref{eq:theta_definition} has been used. Therefore,
\begin{equation}
U^\dagger(\omega)\Gamma(\omega)U(\omega)
=
\operatorname{diag}
\left[
r_+(\omega),r_-(\omega)
\right].
\end{equation}

The corresponding collective jump operators are
\begin{subequations}
\label{eq:Lpm_general_omega}
\begin{align}
\hat L_{+,\omega}
=
\sqrt{r_+(\omega)}
\left[
\cos\theta_\omega\hat A_1(\omega)
\right.
\
\left.
+
e^{i\phi_\omega}
\sin\theta_\omega\hat A_2(\omega)
\right],
\\
\hat L_{-,\omega}
=
\sqrt{r_-(\omega)}
\left[
-\sin\theta_\omega\hat A_1(\omega)
\right.
\
\left.
+
e^{i\phi_\omega}
\cos\theta_\omega\hat A_2(\omega)
\right].
\end{align}
\end{subequations}
Thus, $\theta_\omega$ describes the rotation from the local coupling operators to the eigenchannels of the correlated dissipation. According to Eq. \eqref{eq:A_omega_two_qubits}, the frequencies $\omega_q$ and $-\omega_q$ identify the emission and absorption channels, respectively. For equal local rates, $\Gamma_{11}(\omega)=\Gamma_{22}(\omega)\equiv\Gamma_0(\omega)$, one has $\theta_\omega=\pi/4$, and the eigenvalues reduce to $r_\pm(\omega)=\Gamma_0(\omega)\pm|\Gamma_{12}(\omega)|$.
The jump operators are therefore the symmetric and antisymmetric combinations
\begin{subequations}
\label{eq:channels_sym_rates}
\begin{align}
\hat L_{+,\omega}&=\sqrt{\frac{r_+(\omega)}{2}}\left[\hat A_1(\omega)+e^{i\phi_\omega}\hat A_2(\omega)\right],
\\
\hat L_{-,\omega}&=\sqrt{\frac{r_-(\omega)}{2}}\left[-\hat A_1(\omega)+e^{i\phi_\omega}\hat A_2(\omega)\right].
\end{align}
\end{subequations}

In the maximally correlated rank-one case, $|\Gamma_{12}(\omega)|=\Gamma_0(\omega)$, and hence $r_+(\omega)=2\Gamma_0(\omega), \qquad r_-(\omega)=0$. Only the bright collective channel $\hat L_{+,\omega}$ contributes to the dissipator, while the orthogonal combination $\hat L_{-,\omega}$ identifies the dark channel.

\begin{figure}
    \centering
    \includegraphics[width=\linewidth]{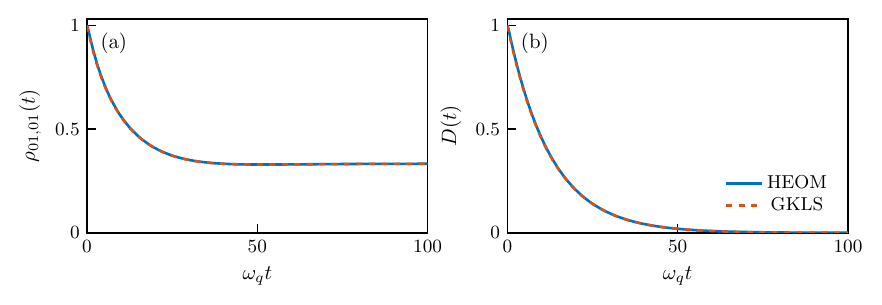}
    \caption{Comparison between the HEOM approach (solid blue lines) and the secular GKLS master equation (dashed orange lines) for two resonant qubits coupled to maximally correlated Drude--Lorentz bath channels. (a) Population $\rho_{01,01}(t)$ for the initial state $\rho_S(0)=\lvert01\rangle\langle01\rvert$. (b) Trace distance $D(t)$, evaluated independently for $\mathrm{HEOM}$ and $\mathrm{GKLS}$. The parameters are $\lambda/\omega_q=0.01$, $\gamma/\omega_q=8$, and $k_BT/\hbar\omega_q=8$. The HEOM calculation uses hierarchy depth $N_C=3$. The correlation function was generated by 15000 Matsubara terms and then fitted with $N_R=3$ real and $N_I=1$ imaginary exponential terms using the least-squares method.}
    \label{fig:lindblad_validation}
\end{figure}

Figure \ref{fig:lindblad_validation} shows an excellent agreement between the HEOM and GKLS dynamics in the weak-coupling regime. In panel (a), both approaches predict the same relaxation of $\rho_{01,01}(t)$ toward the stationary value $\rho_{01,01}\simeq 0.33$. The largest discrepancy between the two curves is approximately $8.4\times10^{-3}$ and occurs during the initial transient, where the finite bath-correlation time retained by the HEOM produces small corrections to the Markovian evolution. Panel (b) shows that the trace distance between the trajectories initialized in $\ket{01}$ and $\ket{10}$ defined as
\begin{equation}
    D(t)=\frac{1}{2}\lVert\rho_{\ket{01}}(t)-\rho_{\ket{10}}(t)\rVert_1.
\end{equation}

$D(t)$ decreases monotonically in both descriptions, without revivals for the selected pair of initial states. The maximum difference between the HEOM and GKLS trace distances remains below $8.3\times10^{-3}$. The agreement of both the transient dynamics and the long-time behavior confirms that the HEOM implementation correctly reproduces the weak-coupling Markovian limit for a rank-one correlated bath.

\subsection{Exactly solvable pure-dephasing benchmark}\label{sec:unruh_validation}

As a second validation, we consider the exactly solvable pure-dephasing limit of the same two-channel system--bath model. We retain the system Hamiltonian of Eq. \eqref{eq:H_sys} and set $\hat Q_\alpha=\hat\sigma_z^{(\alpha)}$. Since $[\hat H_S,\hat H_I]=0$, the populations are constant, whereas the coherences undergo pure dephasing. Denoting the computational basis states by $\ket{\mathbf n}=\ket{n_1n_2}$, with $n_\alpha\in\{0,1\}$, we introduce
\begin{equation}
z_{n_\alpha}=2n_\alpha-1,
\qquad
\hat\sigma_z^{(\alpha)}\ket{\mathbf n} = z_{n_\alpha}\ket{\mathbf n}.
\label{eq:unruh_sigma_z_eigenvalues}
\end{equation}
Thus, $z_{n_\alpha}=-1$ when the $\alpha$-th qubit is in $\ket{0}$ and $z_{n_\alpha}=+1$ when it is in $\ket{1}$.
It is convenient to collect these eigenvalues in the vector
\begin{equation}
\mathbf z_{\mathbf n} = \left(
z_{n_1},z_{n_2} \right).
\label{eq:unruh_coupling_eigenvalue_vector}
\end{equation}
The energy of the computational state $\ket{\mathbf n}$ is
\begin{equation}
E_{\mathbf n} = \frac{\hbar}{2} \left( \omega_1z_{n_1} + \omega_2z_{n_2} \right),
\label{eq:unruh_system_energies}
\end{equation}
and an analogous expression holds for $E_{\mathbf m}$.

Since the interaction is diagonal in the same basis as $\hat H_S$, the bath Hamiltonian conditioned on the system state $\ket{\mathbf n}$ is
\begin{equation}
\hat H_B^{(\mathbf n)} = \hat H_B + \hbar\sum_{\alpha=1}^{2} z_{n_\alpha}\hat B_\alpha.
\label{eq:unruh_conditional_bath_hamiltonian}
\end{equation}
The total propagator can consequently be written as
\begin{equation}
\hat U(t) = \sum_{\mathbf n} e^{-iE_{\mathbf n}t/\hbar} \ket{\mathbf n}\bra{\mathbf n} \otimes
\hat U_{\mathbf n}(t),
\label{eq:unruh_conditional_propagator}
\end{equation}
with $\hat U_{\mathbf n}(t) = \exp\left[ -\frac{i}{\hbar} \hat H_B^{(\mathbf n)}t \right]$.

Taking the matrix element between $\ket{\mathbf n}$ and $\ket{\mathbf m}$ and tracing over the bath gives
\begin{equation}
\rho_{\mathbf n,\mathbf m}(t) = e^{-i(E_{\mathbf n}-E_{\mathbf m})t/\hbar} F_{\mathbf n\mathbf m}(t) \rho_{\mathbf n,\mathbf m}(0),
\label{eq:unruh_reduced_coherence}
\end{equation}

where $F_{\mathbf{n}\mathbf{m}}(t)=\operatorname{Tr}_B\left[U_{\mathbf{n}}(t)\rho_BU_{\mathbf{m}}^\dagger(t)\right]$ is the decoherence function.
For a thermal Gaussian bath, the decoherence function takes the form
\begin{equation}
F_{\mathbf n,\mathbf m}(t) = \exp\left[ -\chi_{\mathbf n,\mathbf m}(t)
+i\Phi_{\mathbf n,\mathbf m}(t) \right],
\label{eq:unruh_influence_function}
\end{equation}
where $\Phi_{\mathbf n,\mathbf m}(t)$ is the bath-induced phase and
\begin{equation}
\chi_{\mathbf n,\mathbf m}(t) = \sum_{\alpha,\beta=1}^{2} \Delta z_\alpha \Delta z_\beta \Lambda_{\alpha\beta}(t),
\label{eq:unruh_dephasing_exponent}
\end{equation}

with $\Delta z_\alpha = z_{n_\alpha}-z_{m_\alpha}$.
The dephasing kernels are determined by the spectral density matrix according to
\begin{equation}
\Lambda_{\alpha\beta}(t) = \int_0^\infty\frac{d\omega}{\pi} \frac{\operatorname{Re}J_{\alpha\beta}(\omega)} {\omega^2}\left[1-\cos(\omega t)\right]\coth\left(\frac{\beta\hbar\omega}{2}\right).
\label{eq:unruh_dephasing_kernel}
\end{equation}

\begin{figure}[t]
\centering
\includegraphics[width=\linewidth]{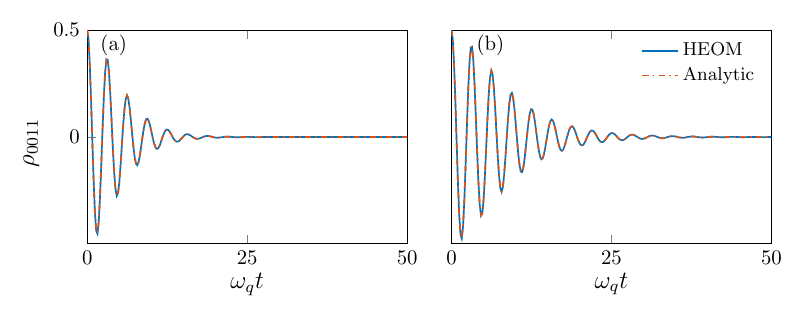}
\caption{Real part of the coherence $\rho_{00,11}(t)$ for the initial Bell state $\ket{\Phi^+}$. The HEOM solution (solid blue line) is compared with the analytic result (dash-dotted orange line) for (a) the rank-one common bath of Eq. \eqref{eq:unruh_common_spectrum} and (b) the independent baths of Eq. \eqref{eq:unruh_independent_spectrum}. The parameters are $\lambda/\omega_q=0.01$, $\gamma/\omega_q=0.2$, and $k_BT/(\hbar\omega_q)=0.2$.  The correlation function was generated by 15000 Matsubara terms and then fitted with $N_R=3$ real and $N_I=1$ imaginary exponential terms using a least-squares method.}
\label{fig:unruh_validation}
\end{figure}

We compare the analytical and HEOM dynamics for two different choices of the spectral density matrix:
\begin{subequations}
\label{eq:unruh_spectral_matrices}
\begin{align}
\mathbf J_{\mathrm{com}}(\omega)
&=J_{\mathrm{DL}}(\omega)
\begin{pmatrix}
1 & 1\\
1 & 1
\end{pmatrix},
\label{eq:unruh_common_spectrum}
\\
\mathbf J_{\mathrm{ind}}(\omega)
&=J_{\mathrm{DL}}(\omega)
\begin{pmatrix}
1 & 0\\
0 & 1
\end{pmatrix}.
\label{eq:unruh_independent_spectrum}
\end{align}
\end{subequations}
Equation \eqref{eq:unruh_common_spectrum} describes a fully correlated common bath and has rank one, whereas Eq. \eqref{eq:unruh_independent_spectrum} describes two independent baths.

For the initial Bell state $\ket{\Phi^+}$, the coherence of interest connects the states $\mathbf n=(0,0)$ and $\mathbf m=(1,1)$, whose coupling-eigenvalue vectors are $\mathbf z_{(0,0)}=(-1,-1)$ and $\mathbf z_{(1,1)}=(+1,+1)$. Consequently, $\boldsymbol{\Delta z} = \mathbf z_{(0,0)} - \mathbf z_{(1,1)} = (-2,-2)$.
Moreover, the two conditional bath displacements have opposite signs and equal squared amplitudes. Hence, the bath-induced phase vanishes for this coherence, $\Phi_{00,11}(t)=0$.
Applying Eq. \eqref{eq:unruh_dephasing_kernel}, the coherences are
\begin{subequations}
\label{eq:unruh_complex_coherences}
\begin{align}
\rho_{00,11}^{\mathrm{com}}(t) &=\frac{1}{2}e^{-16\Lambda(t)} \left[\cos(2\omega_qt)
+i\sin(2\omega_qt) \right],
\label{eq:unruh_common_complex_coherence}
\\
\rho_{00,11}^{\mathrm{ind}}(t) &= \frac{1}{2}e^{-8\Lambda(t)}\left[\cos(2\omega_qt) +i\sin(2\omega_qt) \right].
\label{eq:unruh_independent_complex_coherence}
\end{align}
\end{subequations}

Figure \ref{fig:unruh_validation} compares these analytical expressions with the HEOM dynamics obtained from the exponential decomposition in Eq. \eqref{eq:Corr_func_decomposition}. The agreement in panel (a) validates the treatment of a rank-one correlated bath, including its off-diagonal correlation functions, while panel (b) verifies the independent-bath limit. In the common-bath case, the local and cross-correlation terms add constructively, resulting in faster decay than for the independent baths, where only the two local contributions remain.
Together with the GKLS comparison of Sec. \ref{sec:GKLS}, this benchmark tests the HEOM implementation in both the weak-coupling Markovian regime and an exactly solvable non-Markovian regime.

\section{Weak-coupling stationary reference and long-time limit}
\label{app:StatRef}

For any fixed $\delta_\gamma>0$, suppose that the reduced dynamics relaxes to a unique thermal stationary state. In the weak-coupling limit, this state reduces to the bare Gibbs state $\hat\rho_\beta=e^{-\beta\hat H_S}/Z$ \cite{cresser2021meanforce}. For resonant, noninteracting qubits, $\hat\rho_\beta=\hat\rho_1\otimes\hat\rho_2$, its singlet population is
\begin{equation}
P_D^{(\beta,0)} = \langle\Psi^-|\hat\rho_\beta|\Psi^-\rangle = p_0p_1 = \frac{1}{4} \operatorname{sech}^{2}\!\left( \frac{\beta\hbar\omega_q}{2} \right).
\label{eq:gibbs_singlet_population}
\end{equation}
The corresponding weak-coupling stationary leakage is therefore
\begin{equation}
P_{\mathrm{leak}}^{(\beta,0)} = 1- \frac{1}{4} \operatorname{sech}^{2}\!\left( \frac{\beta\hbar\omega_q}{2} \right),
\qquad
\frac{3}{4} \le P_{\mathrm{leak}}^{(\beta,0)} \le 1.
\label{eq:gibbs_stationary_leakage}
\end{equation}
At $k_BT/(\hbar\omega_q)=0.2$, this gives $P_D^{(\beta,0)}=6.65\times10^{-3}$ and $P_{\mathrm{leak}}^{(\beta,0)}=0.9934$.
Equation~\eqref{eq:gibbs_stationary_leakage} is a weak-coupling stationary reference, not an upper bound on the transient leakage. At finite coupling, the reduced equilibrium state is instead the mean-force Gibbs state and generally depends on the full spectral density.

The limit $\delta_\gamma\to0$ is singular. At $\delta_\gamma=0$, the singlet is exactly protected and $P_{\mathrm{leak}}(t)\equiv0$. By contrast, if every fixed $\delta_\gamma>0$ produces thermalization, then, to leading order in the coupling,
\begin{multline*}
\lim_{\delta_\gamma\to0^+} \lim_{t\to\infty}P_{\mathrm{leak}}(t,\delta_\gamma) = P_{\mathrm{leak}}^{(\beta,0)} \neq 0= 
\\
\lim_{t\to\infty} \lim_{\delta_\gamma\to0^+}P_{\mathrm{leak}}(t,\delta_\gamma).
\end{multline*}
Thus, weak twisting primarily controls the relaxation timescale rather than the leading-order stationary state. If the long-time dynamics is governed by a single slow rate, $\Gamma(\delta_\gamma)=\Gamma_2\delta_\gamma^2 + \mathcal O(\delta_\gamma^4)$, then
\begin{equation}
P_{\mathrm{leak}}(t,\delta_\gamma) \simeq P_{\mathrm{leak}}^\infty \left[ 1-e^{-\Gamma(\delta_\gamma)t} \right].
\end{equation}
The fixed-time quadratic law follows in the regime $\Gamma_2t\delta_\gamma^2\ll1$. At later times, the leakage saturates,
so its dependence on $\delta_\gamma$ is no longer quadratic, even when the relaxation rate itself retains a quadratic weak-twisting onset.

\end{document}